\documentclass[10pt,twocolumn,english,pre,final,byrevtex,showpacs,secnumarabic]{revtex4}
\usepackage[T1]{fontenc}
\usepackage[latin1]{inputenc}
\usepackage{amsmath}
\usepackage{amsfonts}
\usepackage{amssymb}
\usepackage{graphicx}
\usepackage{babel}



\begin{document}

\title{Mean crossover functions for uniaxial 3D Ising-like systems}

\author{Yves Garrabos}

\affiliation{Equipe du Supercritique pour l'Environnement, les Matériaux
et l'Espace, Institut de Chimie de la Matière Condensée
de Bordeaux - UPR 9048, Centre National de la Recherche Scientifique
- Université Bordeaux I - 87, avenue du Docteur Schweitzer, F
33608 PESSAC Cedex, France}

\author{Claude Bervillier}

\affiliation{Laboratoire de Mathématiques et Physique Théorique, CNRS
UMR 6083, Université de Tours, Parc de Gandmont, 37200 Tours,
France.}


\begin{abstract}
We give simple expressions for the mean of the max and min bounds
of the critical-to-classical crossover functions previously calculated
{[}Bagnuls and Bervillier, Phys. Rev. E \textbf{65}, 066132 (2002){]}
within the massive renormalization scheme of the $\Phi_{d}^{4}\left(n\right)$
model in three dimensions ($d=3$) and scalar order parameter ($n=1$)
of the Ising-like universality class. Our main motivation is to get
efficient theoretical expressions to coherently account for many measurements
performed in systems where the approach to the critical point is limited
but yield data which are still reproducible by the $\Phi_{d}^{4}\left(n\right)$
model (like in the subclass of one-component fluids). 
\end{abstract}

\date{05 April 2006}

\maketitle

\section{Introduction}

The universal features of the three-dimensional (3D) Ising-like systems
close to their critical points are now well-established by the renormalization
group (RG) approach \cite{Zinn-Justin2002}. In this theoretical context
the Ising-like universality is attached to the existence of a unique
non-trivial fixed point (the Wilson-Fisher fixed point \cite{Wilson-Fisher1972}
noted W-FP in the following) which any Hamiltonian representation
of an actual system at criticality is driven to under the action of
the renormalization transformations \cite{Wilson-Kogut1974}.

Less known is the existence of theoretical expressions, obtained using
perturbative field theory (FT) techniques \cite{Bagnuls1984a,Bagnuls1985,Bagnuls1987,Schloms1989,Schloms1990},
which are used to interpolate between the critical (non-classical)
behavior (controlled by the W-FP) and a classical behavior (controlled
by the Gaussian fixed point, noted G-FP in the following). Such theoretical
expressions are customarily named classical-to-critical crossover
functions.

Actual systems undergoing a second order phase transition also display
a kind of {}``classical''-to-critical crossover but it is not of
the same nature as the theoretical one alluded to above. In actual
systems, the {}``classical'' part is only a non-critical part (not
governed by the G-FP) whereas in the theoretical interpolation, the
classical part still belongs to a critical domain (governed by the
G-FP). Moreover, the renormalization procedure of FT is reductive
in the sense that many sources of non-universality are discarded because
they are unessential in the vicinity of the critical point. However,
non-universal features become more and more important as one moves
away from the critical point. Finally, these non-universal characteristics
are responsible for the distinction between the nonasymptotic critical
behavior and the asymptotic critical behavior.

Generally, one cannot expect to observe an agreement between the theoretical
crossover functions and the experimental data in a wide domain ranging
from the close vicinity of the critical point down to a state far
away from it. Nevertheless, exceptions may exist as in the case of
the {}``subclass'' of one-component fluids.

It has been shown \cite{Garrabos1982,Garrabos1985,Garrabos1986,Garrabos2002,Garrabos2006}
that, despite their wide range of numerical values, the one component
fluid data collected in the literature regarding a given property
(susceptibility, order parameter density, correlation length, etc.)
can be reduced to a unique scaling curve (also called master curve)
over a wide critical domain. Now, it is well known that this scaling
curve can be well reproduced by the classical-to-critical crossover
functions of FT \cite{Garrabos2006}. However, an accurate understanding
of the range of its validity is lacking. In the present paper, we
provide modified versions (named \emph{{}``mean crossover functions''})
of the theoretical forms given in references \cite{Bagnuls1985,Bagnuls1987,Bagnuls2002}.
A careful attention is given to the characterization of the Ising
like preasymtotic domain where the number and the nature of adjustable
parameters can be easily controlled. A forthcoming paper \cite{Garrabos2006a}
will illustrate the application of the mean crossover functions to
provide a precise criterion for estimating the effective extension
of the critical asymptotic domain of the fluid subclass, a key point
for future developments of a \emph{complete} equation of state (e.o.s.)
for fluids.

Actually, since it is difficult to approach accurately the critical
point both experimentally and theoretically, the effective (observed)
scaling domain is limited towards the critical point whereas, in the
opposite direction, the scaling curve stops when the systems are no
longer critical and before they display any kind of {}``classical
behavior''. Here classical behavior means the {}``critical'' behavior
controlled by the G-FP. Consequently, most of the useful data on a
system near criticality belong to an intermediate (nonasymptotic)
regime where it is not valid to reproduce the data with the currently
used expansion limited to a pure power law eventually corrected by
one Wegner's term \cite{Wegner1972} (the validity of such a form
implies the close vicinity of the critical point).

For development of an e.o.s., it is of great importance to have theoretical
expressions which reproduce the phenomenologically observed scaling
curves and which allow one to situate unambiguously any data set with
respect to the actual (system dependent) field distance to the critical
point. It must be stressed that, in this view, the aim is not to test
the most refined theoretical hypotheses, or the most precise estimates
of the universal numbers associated with the asymptotic or pre-asymptotic
critical behavior. In addition, our scheme only applies to the primary
critical paths along the critical isochore in the homogeneous and
the non homogeneous domain, without any supplementary consideration
on the roles played by other nonuniversal scales associated with the
eventual effects such as non symmetry \cite{Anisimov2000} and mixing
\cite{Kim2003} of the scaling field variables very close to the critical
point.

>From a pioneering study \cite{Bagnuls1984b} of the available data
on xenon \cite{Guttinger1981,Edwards1968}, it is known that the classical-to-critical
crossover functions obtained from the $\Phi_{d}^{4}\left(n\right)$
model (with $d=3$ and $n=1$) fit the one-component fluid results,
introducing explicit determination of the fluid-dependent \emph{scale
factor} for the relative temperature field. More recently, a similar
study \cite{Zhong2004} of several measurements near the liquid-gas
critical point of $^{3}$He \cite{Zhong2003}, has provided an equivalent
conclusion, demonstrating that, $^{3}$He as a simple fluid, is also
situated very close to a renormalized trajectory that links the G-FP
to the W-FP. However, the authors of Ref \cite{Bagnuls1984a,Bagnuls1985,Bagnuls1987,Bagnuls2002}
have given a practical (max or min) form of their effective functions
which is better adapted to the eventual experimental test of the {}``best''
theoretical estimates of the asymptotic universal quantities (attached
to the close vicinity of the critical point), rather than to an unambiguous
characterization of a nonasymptotic critical domain of a given subclass
of systems. In fact, accounting for the error estimates attached to
the universal critical quantities under two \emph{max} and \emph{min}
bounded sets of functions prevents the characterization of the scaling
curve clearly which, of course, is currently accessible experimentally
at some non-small distance (to be determined) to the critical point
but also is sometimes accompanied by a relatively poor accuracy in
the measurement. Moreover, due to the decrease of the theoretical
error as the distance to the critical point increases, the mean values
of the bounded functions are sufficient to characterize the scaling
curves.

Our aim in the present paper is to determine in-between controlled
functions that are better adapted for unambiguous determination (within
a current experimental mean accuracy) of the characteristics of the
scaling curves of the subclass of one-component fluids. An important
point to note a posteriori is the unchanged value of the temperature
scale factor determined in \cite{Bagnuls1984b}. However the uncertainty
is considerably reduced showing that it was essentially generated
by the theoretical uncertainties of the universal values calculated
in the asymptotic regime. That demonstrates the need to provide mean
classical-to-critical crossover functions which, by construction,
have a well defined single asymptotic limit (\textit{i.e. {}``Ising-like-well-defined''}).
We will then be able to determine significative values of the scale
factors which characterizes the scaling behavior occuring in the intermediate
critical domain.

The theoretical crossover functions which we are interested in have
been derived from a massive renormalization (MR) scheme applied to
the $\Phi_{d}^{4}\left(n\right)$ model. The MR scheme has been initially
developed in references \cite{Bagnuls1984a,Bagnuls1985,Bagnuls1987},
hereafter referenced MR6, using the Borel resummation technique based
on the results of the sixth-loop series \cite{Nickel1977}. They have
been recently revisited in reference \cite{Bagnuls2002}, hereafter
referenced MR7, to account for an extension to the seventh-loop series
\cite{Murray-Nickel1991}. We do not consider here the crossover functions
determined by Dohm and coworkers \cite{Schloms1989,Schloms1990} who
have used another renormalization scheme (minimal) within which the
known series are shorter than in the MR scheme and have provided their
crossover functions under implicit forms.

The paper is organized as follows.

In Section 2, we introduce the main characteristics of the MR scheme
and of the crossover functions. Special attention is given to the
non-universal nature of the adjustable parameters that are introduced
in the max and min theoretical functions for fitting experimental
data.

In Section 3, the mean functions are determined relying on the properties
of the theoretical functions in the two limiting \emph{3D Ising-like}
and \emph{mean field-like} descriptions, respectively close to the
W-FP and G-FP. Such descriptions correspond to the pre-asymptotic
domains (PAD) near each fixed point where a Wegner expansion restricted
to two terms (leading and first confluent terms) is valid. The \emph{Ising-like}
PAD includes the correlations between parameters due to the error-bar
determination of the exponents and amplitude combinations very close
to the W-FP. The addition of the equivalent \emph{mean field} PAD
description very close to the G-FP leads to a mean crossover function
with a limited number of calculated parameters for this function (three
in the selected example). A well-controlled form of any mean crossover
function can be obtained in a similar manner, i.e. three calculated
parameters for each function. In such a situation, the theoretical
crossover forms are obtained for a unique value of one parameter among
the three. This parameter acts then as a relative \emph{sensor} to
estimate the dominant nature, either (Ising-like) \emph{critical},
or (mean field-like) \emph{classical}, of the calculated crossover.
Using this sensor, we propose an explicit criterion to measure the
extension of the Ising-like PAD along the critical isochore where
a four parameter characterization of each system is well-understood
in scaling nature. In section 4, we provide a conclusion.

In Appendix A, assuming knowledge of the critical temperature of the
system, we illustrate the three-adjustable-parameter characterization
of the comparison with experimental data. The emphasis is on the role
of a characteristic microscopic length scale, the reminiscence of
which is carried by the dimension of the {}``bare'' $\phi^{4}$-coupling
$g_{0}$ {[}see Eq. (\ref{Phi4 hamilton (1)}) below{]}.

Appendix B gives some details on the derivation of the mean crossover
function for the particular case of the order parameter in the heterogeneous
domain, with a view to better account for large (theoretical and experimental)
uncertainties in the determination of its corresponding first confluent
amplitude.

\section{The massive renormalization scheme}

\subsection{The model \label{model}}

The calculations of the crossover functions in the $\Phi_{d}^{4}$
model rely upon the renormalization program of perturbation FT. This
scheme makes the perturbative expansion of the correlation functions
free of ultra-violet divergencies. Before renormalization, the dimensionless
Hamiltonian (a true hamiltonian divided by $k_{B}T$, with $k_{B}$
the Boltzman constant) of the actual system reads: \begin{align}
\mathcal{H}=\int d^{d}x & \left\{ \frac{1}{2}\left[\left(\bigtriangledown\overrightarrow{\Phi_{0}}\right)^{2}+r_{0}\left(\overrightarrow{\Phi_{0}}\right)^{2}\right]+\frac{g_{0}}{4!}\left[\left(\overrightarrow{\Phi_{0}}\right)^{2}\right]^{2}\right.\nonumber \\
 & \left.+\vec{h}.\overrightarrow{\Phi_{0}}\right\} \label{Phi4 hamilton (1)}\end{align}
 in which $d$ is the space dimension and $\vec{h}$ and $\overrightarrow{\Phi_{0}}$
are vectors of dimension $n$. $\vec{h}$ is the (magnetic-like) ordering
field vector. $\overrightarrow{\Phi_{0}}$ is the spin-like vector.
The ordering field vector $\vec{h}$ is not renormalized and so will
no longer be considered explicitly in the following (except when required).
The coupling field $r_{0}$ and the coupling constant $g_{0}$ are
the bare (physical) parameters, which are two system-dependent quantities
characterizing the critical point location of the physical system.

For the sake of simplicity, we limit ourselves in the following discussion
to the scalar case $n=1$.

At this stage, it is worthwhile to state that the dimension of any
quantity appearing in $\mathcal{H}$ is expressed in terms of only
one inverse length unit: a wave-vector cutoff, $\Lambda$ (or the
inverse of a lattice spacing for magnetic solid systems for example).
Since $\mathcal{H}$ is dimensionless, a simple evaluation of Eq.
(\ref{Phi4 hamilton (1)}) shows that the dimension (in unit $\Lambda$)
of $\Phi_{0}$ is $\frac{d-2}{2}$, while that of $r_{0}$ is $2$,
and that of $g_{0}$ is $4-d$.

When $d=4$, the perturbative expansion in powers of the dimensionless
coupling constant $g_{0}$ involves ultra-violet divergencies which
are removed by redefining the three initial Hamiltonian (bare) parameters
$\left\{ \Phi_{0},r_{0},g_{0}\right\} $ into new (renormalized) ones
$\left\{ \Phi,\overline{m},g\right\} $, via the following relations:
\begin{equation}
r_{0}=\overline{m}^{2}+\delta\overline{m}^{2}\label{mass renorm (2)}\end{equation}
\begin{equation}
\Phi_{0}=\left[Z_{3}\left(g\right)\right]^{\frac{1}{2}}\Phi\label{op renorm (3)}\end{equation}
\begin{equation}
g_{0}=\left(\overline{m}\right)^{4-d}g\frac{Z_{1}\left(g\right)}{\left[Z_{3}\left(g\right)\right]^{2}}\label{coupling cte renorm (4)}\end{equation}
 with subtraction conditions which, in four dimensions ($d=4$), are
required to make the renormalized perturbative expansion in powers
of $g$ finite.

In three dimensions, however, only the mass renormalization is needed
to have a finite theory. Consequently, with the introduction of $\overline{m}$
instead of $r_{0}$, the renormalization functions, like $Z_{1}$
and $Z_{3}$, even expressed in terms of the {}``bare'' coupling
$g_{0}$ and without ultra-violet regulator $\Lambda$ (set to infinity),
are well defined functions. Introducing the notation $\Gamma_{0}^{\left(L,N\right)}\left(\left\{ q,p\right\} ;\overline{m},g_{0},d\right)$
for the Fourier transforms of the bare $N$-point vertex functions
with $L$ {}``insertions'' of the squared field $\Phi_{0}^{2}$,
the following definitions stand (for $d<4$)\begin{align*}
Z_{3}^{-1} & =\frac{\partial}{\partial p^{2}}\Gamma_{0}^{\left(0,2\right)}\left(p;\overline{m},g_{0},d\right)\\
Z_{1}^{-1} & =\frac{\Gamma_{0}^{\left(0,4\right)}\left(\left\{ 0\right\} ;\overline{m},g_{0},d\right)}{g_{0}}\\
Z_{2}^{-1} & =\Gamma_{0}^{\left(1,2\right)}\left(\left\{ 0\right\} ;\overline{m},g_{0},d\right)\end{align*}
 in which $Z_{2}$ is a renormalization function which restores the
linear measure of the distance to the critical temperature $T_{c}$,
originally defined by the bare parameter $r_{0}$ and which has been
lost when introducing the renormalized mass $\overline{m}$ {[}via
Eq. (\ref{mass renorm (2)}) and a condition, not written here, which
defines $\overline{m}$ as the inverse correlation length{]}.

Since $\Lambda$ has been eliminated in the renormalization process,
$\overline{m}$ plays the role of the effective wave-vector scale
of reference and could be used as the unit to express the dimension
of any dimensioned quantity. However, because $\overline{m}$ (the
inverse of the correlation length) vanishes at the critical point
it is preferable to use $g_{0}$ which, for $d\neq4$, is a dimensioned
constant at the critical temperature {[}see Eq. (\ref{gzero (10)})
below{]}. Hence, $\left(g_{0}\right)^{\frac{1}{4-d}}$ will be substituted
to $\Lambda$ to play the role of the wave-vector unit (see below).
Notice that in three dimensions $g_{0}$ has exactly the dimension
of $\Lambda$ (i.e. of the inverse of a length) which is very convenient
but not essential.

The dimensionless renormalization functions $Z_{i}$ have been calculated
up to sixth order \cite{Nickel1977} and then partly up to seventh
order \cite{Murray-Nickel1991} in powers of $g$. These series have
been summed to estimate the critical exponents with great accuracy
\cite{LeGuillou-Zinn1980,Guida1998} and also to determine nonasymptotic
critical functions in the homogeneous phase \cite{Bagnuls1984a,Bagnuls1985}.
With the calculation of supplementary integrals \cite{Bagnuls1987},
the calculations have been exended to the inhomogeneous phase up to
fifth order allowing the determination of the nonasymptotic critical
functions \cite{Bagnuls1987} in this phase and of the equation of
state \cite{Guida1997,Guida1998}. The calculations of the nonasymptotic
critical functions have been revisited \cite{Bagnuls2002} in order
to account for the most recent estimates of the asymptotic universal
critical quantities \cite{Guida1998} and also to provide complete
classical-to-critical crossover forms \cite{Bagnuls2002} which we
are presently interested in.

\subsection{The crossover functions}

In the perturbative framework, the renormalized coupling $g$ may
take on any value in the range $[0,g^{\ast}]$ where $g^{\ast}$ is
its W-FP value (a value which may be estimated in three dimensions
by looking at the nontrivial zero of some series) and $0$ is its
G-FP value. The crossover functions have been obtained by resumming
the series of a physical property $P$ of interest such as the correlation
length $\ell\left(g\right)$ {[}$=\left(\overline{m}\right)^{-1}${]}
or the susceptibily $\chi\left(g\right)$, for a discretized variation
of $g$ in this range. An expression for the critical scaling field
$t\left(g\right)$ (which {}``measures'' the physical distance to
the critical point) is obtained via an integration of the resummed
series of $Z_{2}\left(g\right)$ (see ref. \cite{Bagnuls1985} for
example). Hence the variations of $\ell$ or $\chi$, in terms of
$t$, are primarily obtained implicitly via the dummy parameter $g$.
Explicit functions of $t$ representing $\ell$ or $\chi$ are then
obtained by fitting ad hoc forms to their discretized evolutions in
the range $]0,g^{\ast}[$.

It is to be noted that, appart from $r_{0}$ which is used to determine
$t$ (see below), the only remaining dimensioned parameter of the
$\Phi_{d}^{4}$ model is the bare coupling $g_{0}$ with a dimension
(in length unit) equal to $d-4$ . Then, all the final functions are
reduced by the appropriate powers of $g_{0}$ so as to be dimensionless.

In MR6, contrary to MR7, the entire crossover (corresponding to the
complete range $]0,g^{\ast}[$) had not been published. Also, in MR6,
the error analysis was made independently from that associated with
the estimates of the critical exponents done by Le Guillou and Zinn-Justin
\cite{LeGuillou-Zinn1980}. In MR7 the convergence criteria for the
Borel resummation of the different functions, like $\ell$ or $\chi$,
was chosen such that the \emph{max} and \emph{min} bounds of the resulting
critical exponents agreed as closely as possible to the (revised)
values obtained by Guida and Zinn-Justin \cite{Guida1998}. However,
in doing so, it is more than likely that the error bars have been
over estimated compared to what the resummation method used would
have naturally indicated (following the rules applied in MR6). This
is why the results of MR6 are also of interest to us especially in
the case of the order parameter in the heterogeneous domain for which
the large uncertainties of MR7 are not favourable for accurate fitting
of the experimental data (see appendix B).

In MR7, two tables were presented to give an envelope for each function
$F_{P}\left(t^{\ast}\right)$ representing a (dimensionless) model
property $P$ versus a discretized dimensionless scaling field $t^{\ast}$
{[}defined in Eq. (\ref{tstar (11)}) below{]} over the entire range
$\left]0,\, g^{\ast}\right[$ for the \emph{max} and \emph{min} bounds.
The following expressions $F_{P}\left(t^{\ast}\right)$ were used
to continuously fit the discrete data from each particular table \begin{equation}
F_{P}\left(t^{\ast}\right)=\mathbb{Z}_{P}^{\pm}\left(t^{\ast}\right)^{-e_{P}}\prod_{i=1}^{K}\left(1+X_{P,i}^{\pm}\left(t^{\ast}\right)^{D_{P}^{\pm}\left(t^{\ast}\right)}\right)^{Y_{P,i}^{\pm}}\label{Function MR P (5)}\end{equation}
 with \begin{equation}
D_{P}^{\pm}\left(t^{\ast}\right)=\Delta-1+\frac{S_{P,1}^{\pm}\sqrt{t^{\ast}}+1}{S_{P,2}^{\pm}\sqrt{t^{\ast}}+1}\label{Deff exponent MR P (6)}\end{equation}

In Eq. (\ref{Function MR P (5)}), $3\leq K\leq5$, depending on the
required fit quality for each property $P=\left\{ \left(\chi^{*}\right)^{-1};\left(\ell^{*}\right)^{-1};C^{*};m^{*}\right\} $
{[}the superscript {*} indicates dimensionless quantities, $\chi^{*}$
is the susceptibility; $\ell^{*}$ is the correlation length; $C^{*}$
is the heat capacity; $m^{*}$ is the order parameter in the non-homogeneous
domain (note the distinction with the decorated $\overline{m}$ used
for the renormalized mass){]}. $e_{P}$ and $\Delta$ are the leading
and first confluent universal exponents, respectively. The symbol
$\pm$ indicates the possible homogeneous ($+$) and nonhomogeneous
($-$) domains. All the constants $\mathbb{Z}_{P}^{\pm}$, $X_{P,i}^{\pm}$,
$Y_{P,i}^{\pm}$, and $S_{P,i}^{\pm}$ in Eqs. (\ref{Function MR P (5)})
and (\ref{Deff exponent MR P (6)}) are tabulated in \cite{Bagnuls2002}.
They result from the fitting of eqs. (\ref{Function MR P (5)}, \ref{Deff exponent MR P (6)})
to the theoretical calculations done point by point in the discretized
complete $t^{\ast}=\left\{ \infty,0\right\} $ range corresponding
to the complete range $]0,g^{\ast}[$. The critical behavior of the
specific heat is particular such that, compared to Eq. (\ref{Function MR P (5)}),
it involves an additive critical constant that we note $X_{P,6}^{\pm}$.

\subsection{Physical validity of the functions}

\subsubsection{Analytical corrections discarded}

To compare the theoretical functions to measurements, we must relate
the Hamiltonian parameters ($r_{0}$, $g_{0}$, $h$) to their physical
counterparts. This is done by the (usual) basic assumption that the
bare quantities $\left(g_{0}\text{, }r_{0}\text{, }h\right)$ are
analytical functions of the corresponding physical quantities $T$
and $H$ (in usual notations for magnetic systems) which control the
approach to the actual critical point. Implicitly, we admit that the
Hamiltonian energy is comparable to the physical free energy measured
in unit of $k_{B}T\cong k_{B}T_{c}$ very close to the actual critical
temperature $T_{c}$ of the system.

Assuming that analytical corrections to scaling are negligible, this
introduces two arbitrary scale factors (noted $\vartheta$ and $\psi$
in the following) associated with the two bare fields $r_{0}$ and
$h$, respectively, and one constant inverse length scale $\left(g_{0}\right)^{\frac{1}{4-d}}$
which fixes the dimensionality of the physical variables. This can
be accomplished as follows:

\begin{enumerate}
\item At $h=0$, the bare field $r_{0}=r_{0}\left(T\right)$ must be related
to the actual temperature $T$ of the system, leading to define its
critical value $r_{0c}=r_{0}\left(T_{c}\right)$ from the actual critical
temperature $T_{c}$ of the system. For $T$ close to $T_{c}$, it
becomes possible to relate the bare field difference $r_{0}\left(T\right)-r_{0c}\left(T_{c}\right)$
to the temperature distance $T-T_{c}$ by the following analytical
(linear) approximation\begin{align}
r_{0}-r_{0c}=\mathrm{const\times}\left(T-T_{c}\right)+\mathcal{O}\left[\left(T-T_{c}\right)^{2}\right]\label{rzero vs T-Tc (7)}\end{align}
 In an equivalent manner, at $r_{0}=r_{0c}$, the bare field $h$
must be related to the actual (magnetic field) variable $H$ (in notations
for magnetic system) of the system, leading also to the possible analytical
(linear) approximation\begin{equation}
h=\mathrm{const\times}H+\mathcal{O}\left[H^{2}\right]\label{h vs H (8)}\end{equation}
 for $h$ close to zero. Correlatively, the linearization between
the bare order parameter $m=\left\langle \Phi_{0}\right\rangle $
to the actual (magnetization) density variable $M$ (in notations
for magnetic system) reads\begin{equation}
\begin{array}{cl}
m= & \mathrm{const\times}M\left[\left(T-T_{c}\right),H\right]+\\
 & \mathcal{O}\left[\left(T-T_{c}\right)^{2},H^{2},H\times\left(T-T_{c}\right)\right]\end{array}\label{m vs M (9)}\end{equation}

where the two constant prefactors of Eqs. (\ref{h vs H (8)}) and
(\ref{m vs M (9)}) are interrelated by thermodynamic considerations
attached to the conjugated variables $M$ and $H$ (or $m$ and $h$
equivalently).

Finally, at the above linearized order of the relations between the
bare fields and the physical fields, close to the critical point defined
by $T-T_{c}=0$ (or $r_{0}-r_{0c}=0$) and $H=0$ (or $h=0$), we
can introduce one finite value of the coupling constant such as,\begin{align}
g_{0} & =\mathrm{const}+\mathcal{O}\left(T-T_{c}\right)+\mathcal{O}\left[H\right]\label{gzero (10)}\end{align}
 which appears then as a system-dependent critical quantity which
must take the $d-4$ dimension (in length unit). In the Eqs. (\ref{rzero vs T-Tc (7)})
to (\ref{m vs M (9)}), one generally keeps only the leading terms
but one must keep in mind that far away from $T_{c}$, the neglected
second order analytical terms $\mathcal{O}\left[\left(T-T_{c}\right)^{2}\right]$,
$\mathcal{O}\left[H^{2}\right]$, and $\mathcal{O}\left[H\times\left(T-T_{c}\right)\right]$
(within $r_{0}$, $h$ and $m$) could have some importance especially
in approaching the G-FP.

\item With $g_{0}$ defined by Eq. (\ref{gzero (10)}), we can now appropriately
make dimensionless the bare quantity $r_{0}$ by introducing the dimensionless
scaling field $t^{*}$ used in Eqs. (\ref{Function MR P (5)}) and
(\ref{Deff exponent MR P (6)}): \begin{equation}
t^{\ast}=\frac{r_{0}-r_{0c}}{\left(g_{0}\right)^{\frac{2}{4-d}}}\label{tstar (11)}\end{equation}
 For the actual system, it is convenient to define the reduced temperature
distance in units of the critical temperature (already chosen to express
the energy unit of the hamiltonian), leading to the usual notation
of the thermal like field $\Delta\tau^{*}=\frac{T-T_{c}}{T_{c}}$.
Assuming a {}``sufficiently'' small $\Delta\tau^{*}$, $t^{\ast}$
and $\Delta\tau^{*}$ are related by the first dimensionless arbitrary
scale factor $\vartheta$:\begin{equation}
t^{\ast}=\vartheta\times\Delta\tau^{*}\label{theta RG linearization (12)}\end{equation}
 Similar assumptions stand for the second scaling field $h$ which
must be related to $H$ by Eq. (\ref{h vs H (8)}).

A second arbitrary scale factor appears, noted $\psi$, between the
dimensionless field $h^{\ast}=\frac{h}{\left(g_{0}\right)^{\frac{d+2}{2\left(4-d\right)}}}$
and its corresponding dimensionless physical quantity $H^{*}$ (here
in notations for magnetic systems):\begin{equation}
h^{\ast}=\psi\times H^{*}\label{psy RG linearization (13)}\end{equation}
 Correspondingly, the dimensionless order parameter field $m^{\ast}=\left\langle \Phi_{0}^{\ast}\right\rangle =\frac{m}{\left(g_{0}\right)^{\frac{d-2}{2\left(4-d\right)}}}$
is then related to the dimensionless physical quantity $M^{\ast}$
(here in notation for magnetic systems) by \begin{equation}
m^{\ast}=\left(\psi\right)^{-1}\times M^{*}\label{OP RG linearization (14)}\end{equation}
 Finally the {}``length'' $\left(g_{0}\right)^{\frac{1}{d-4}}$
must be related to a microscopic length scale $a$ characteristic
of the actual system as: \begin{equation}
\left(g_{0}\right)^{\frac{1}{d-4}}=u_{0}^{*}\times a\label{microlengthscale (15)}\end{equation}
 where $u_{0}^{*}$ is a dimensionless number similar to $\vartheta$
and $\psi$. Notice that $u_{0}^{*}$ has the characteristic of taking
a value which depends on the (a priori) choice of a system dependent
length scale $a$, not attached to the critical behavior. That characteristic
will be used to determine a subclass of {}``comparable'' systems
{[}i.e. systems having a comparable characteristic length scale $a$
(see below){]}.

As will be shown in Appendix A, it is worthwhile already indicating
here that the consideration of $a$ via Eq. (\ref{microlengthscale (15)})
is not required when the correlation length is considered alone. This
is because the dimensionless correlation length $\ell^{*}$ of the
theoretical model is naturally compared with the experimental measurement
$\xi$ via the following relation: \begin{equation}
\ell^{*}=\left(g_{0}\right)^{\frac{1}{4-d}}\times\xi\label{ellstar (16)}\end{equation}

\end{enumerate}
In Appendix A, limiting our calculations to the $d=3$ case, we illustrate
the way the adjustable parameters are determined where the emphasis
is also on the choice of the microscopic length scale $a$ appearing
in Eq. (\ref{microlengthscale (15)}). When the critical temperature
of the system is known, fitting the asymptotic two-term expansion
of our theoretical crossover functions to the singular behavior of
the experimental quantities, permits unambiguous determinations of
the three system-dependent parameters $\vartheta$, $\psi$, and $g_{0}$.
We then indicate how the dimensionless coupling constant $u_{0}^{*}$
may be used to characterize {}``comparable'' systems which belong
to a same subclass of universality when an explicit length scale unit
($a$) is known for each system. We show that the theoretical crossover
functions then provide an explicit analytical form of the {}``master''
(i.e. unique) singular behavior associated to this subclass and discriminate
the role of the energy and length scale factors for the subclass.

\subsubsection{Non-analytical corrections discarded -- \protect \protect \protect \protect \protect \\
 Some ideas on renormalization}

The renormalized $\Phi_{d}^{4}\left(n\right)$ model of the perturbative
FT is an efficient reduction of a complicated mathematical problem
which originally involves an infinite number of parameters to one
parameter $g$ (the renormalized $\Phi^{4}$ coupling $g$ of section
\ref{model}). In order to better understand the impact of this reduction,
it is necessary to consider the complete and non-perturbative renormalization
group theory developed by Wilson \cite{Wilson-Kogut1974}.

Explaining the use of the renormalization theory in the study of critical
phenomena is out of the scope of the present paper. However, it is
worthwhile to indicate briefly some general ideas which may help someone
to understand the use of our crossover functions in comparison with
experimental data.

The RG theory is a general theory to treat situations where infinitely
many degrees of freedom are correlated, such as near a critical point
where the correlation length $\xi$ diverges. Since in such cases,
we are essentially interested in describing the large distance behavior,
and because of the correlations, one may represent the state of a
system near criticality by means of a dimensionless hamiltonian $\mathcal{H}$
which depends only on a local field $\phi\left(x\right)$ which summarizes,
over a volume of linear size $\Lambda^{-1}$ centered on $x$, only
general properties of the genuine local variable which critically
fluctuates (e.g. the spin variables $S_{i}$ of the Ising system).
Beyond $\phi\left(x\right)$, $\mathcal{H}$ depends also on the relevant
physical parameters like the temperature $T$ and the magnetic field
$H$ (for the sake of simplicity we shall not consider $H$). The
form of $\mathcal{H}$ is quite general provided it satisfies some
required properties of symmetry. For systems which are $O(1)$-symmetric
{[}more precisely $Z_{2}$-symmetric, i.e. invariant under the change
$\phi\rightarrow-\phi$ such as Ising-like systems{]}, then $\mathcal{H}$
must be an even function of $\phi$ (when $H=0$).

In order to concretize a bit the form of $\mathcal{H}$, one may look
at its expansion for small values of $\phi\left(x\right)$, in the
case of the $Z_{2}$-symmetry, it becomes ($a_{0}$ being a term which
is usually discarded in FT) : 
\begin{eqnarray}
	\mathcal{H} & \mathcal{=} & a_0 + \int d^{d}x \left\{ a_1 \left( \nabla \phi \right)^{2} + a_2 \phi^{2} + a_{3} \phi^{4} \right.
	\nonumber \\
 &  & \left.+a_{4}\phi^{6}+a_{5}\phi^{2}\left(\nabla\phi\right)^{2}+\cdots\right\} 
	\label{(17)}
\end{eqnarray}
 in which the expansion in powers of the derivatives is a consequence
of the short range interactions between the original spins and the
coefficients $a_{i}$ depend on $T$. Actually one is free to redefine
the global normalization of the field so as to set $a_{1}=$ constant
(in general one chooses $a_{1}=\frac{1}{2}$) but this is not mandatory.
Furthermore all the dimensions are measured in terms of $\Lambda$,
so that it is convenient to deal with dimensionless quantities in
$\mathcal{H}$. The set $\left\{ a_{i}\left(\frac{T}{T_{c}}\right)\right\} $
characterizes a given physical system near a critical temperature
$T_{c}$. It may be seen as the coordinate of a point in a space $\mathcal{S}$
of infinite dimension.

The problem is that calculation of any critical quantity with $\mathcal{H}$
is very complicated. The reason is that the parameters at hand $\left\{ a_{i}\left(\frac{T}{T_{c}}\right)\right\} $
are attached to the length scale $\Lambda^{-1}$, while the physics
under consideration refers to phenomena that occur at all the length
scales smaller than $\xi$ and greater than $\Lambda^{-1}$. Now because
the ratio $\frac{\xi}{\Lambda^{-1}}$ tends to infinity, there are
an infinity of scales to be accounted for to solve the problem.

Introducing a parameter $s$ such as 
$s\in\left[0,\,\infty\right[$,
the renormalization group transformations precisely construct successive
hamiltonians $\mathcal{H}\left(s\right)$, obtained by integrating
out the degrees of freedom over the scales ranging in $\left[\left(e^{-s}\Lambda\right)^{-1},\Lambda^{-1}\right]$,
and then rescaling back the wave-vector scale (hence $\Lambda^{\prime}=e^{-s}\Lambda\rightarrow\Lambda$).
By construction, the renormalized hamiltonian $\mathcal{H}\left(s\right)$
presents a correlation length $\xi_{s}$ which is reduced, compared
to the $\xi$ of the initial hamiltonian $\mathcal{H}\left(s=0\right)=\mathcal{H}$,
with $\xi_{s}=e^{-s}\xi$, and the effective hamiltonian $\mathcal{H}\left(s\right)$
is no longer critical when $\frac{\xi}{\Lambda^{-1}}\simeq1$. If
the initial hamiltonian is critical (i.e. if $\xi=\infty$), then
$\mathcal{H}\left(s\right)$ reaches a fixed point when $s\rightarrow\infty$.

The evolution of $\mathcal{H}\left(s\right)$ under an infinitesimal
change of $s$ is governed by an equation, called the exact RG equation.
It is a complicated integro-differential equation which may be expanded
in powers of the derivative of the field without losing the non-perturbative
character of the complete theory (for a review see \cite{Bagnuls2001}).

At leading order of the derivative expansion, called the local potential
approximation (LPA), the exact RG equation reduces to an ordinary
differential equation for a function $V$ of one variable $\phi$.
Its relation to the $\Phi_{d}^{4}$ model is better seen if one expands
$V$ in powers of $\phi$. Imposing the property of parity in $\phi$,
it becomes (for $\phi$ close to $0$):\begin{equation}
V\left(\phi,s\right)=V_{0}\left(s\right)+V_{2}\left(s\right)\phi^{2}+V_{4}\left(s\right)\phi^{4}+V_{6}\left(s\right)\phi^{6}+\cdots\label{Wilson-Potential (18)}\end{equation}

The form of Eq. (\ref{Wilson-Potential (18)}) shows the relation
with Eq. (\ref{Phi4 hamilton (1)}) when the derivatives of the field
are neglected (again, for the sake of simplicity, the account of $h$,
which would require the presence of odd powers of $\phi$, is not
considered here). In particular, the initial form chosen for $V$
may be directly related to the bare parameters of the $\Phi_{d}^{4}$
model with, e.g. $\Lambda^{2}V_{2}\left(0\right)=\frac{r_{0}}{2}$,
and $\Lambda^{4-d}V_{4}\left(0\right)=\frac{g_{0}}{4!}$ (and $V_{0}\left(0\right)=V_{6}\left(0\right)=0)$,
they are parameters attached to the microscopic length scale $\Lambda^{-1}$
of the initial hamiltonian ($s=0$). As for the renormalized coupling
$g$ of the perturbative framework, it is not a constant but a running
parameter like $V_{4}\left(s\right)$. To be more precise, one must
look at the flow of $V\left(\phi,s\right)$ solution of the RG equation.

In LPA, the derivatives of the field are neglected but one may account
for all the effects of all the Hamiltonian terms via the evolution
of the complete function $V\left(\phi\right)$ under the action of
the RG transformation.

In three dimensions, a non-trivial fixed point $\mathring{V}\left(\phi\right)$
exists and may be (numerically) determined. One may illustrate this
by writing (as a formal polynomial of $\phi$):\begin{equation}
\mathring{V}\left(\phi\right)=\mathring{V}_{0}+\mathring{V}_{2}\phi^{2}+\mathring{V}_{4}\phi^{4}+\mathring{V}_{6}\phi^{6}+\cdots\label{dimensionless Wilson potential (19)}\end{equation}
 and then calculating the values of $\mathring{V}_{0}$, $\mathring{V}_{2}$,
$\mathring{V}_{4}$, $\mathring{V}_{6}$, etc. As a consequence, one
may situate the fixed point in the formal space $\mathcal{S}$ truncated
to the set of couplings $\left\{ V_{0},V_{2},V_{4},V_{6},\cdots\right\} $
of infinite dimension. To visualize the evolution of a RG trajectory
of $V\left(\phi,s\right)$ (the evolution of $V\left(\phi\right)$
under a continuous RG transformation controled by the parameter $s\in\left[0,\,\infty\right[$
in $\mathcal{S}$), one may project it in the plane $\left\{ V_{4},V_{6}\right\} $,
for example.

We may choose to integrate the differential equation with an initial
simple form for $V\left(\phi,s=0\right)$ like, for example:\begin{equation}
V\left(\phi,0\right)=V_{2}\left(0\right)\phi^{2}+V_{4}\left(0\right)\phi^{4}+V_{6}\left(0\right)\phi^{6}\label{WilPotInit (20)}\end{equation}
 in which $V_{2}\left(0\right)$, $V_{4}\left(0\right)$ and $V_{6}\left(0\right)$
are numbers representing the initial values of $V\left(s\right)$.
As soon as $s\neq0$, then $V\left(\phi,s\right)$ contains all powers
of $\phi$.

To approach the fixed point $\mathring{V}\left(\phi\right)$ starting
with (\ref{WilPotInit (20)}), one must adjust one of the three initial
parameters, e.g. $V_{2}(0)$, to a particular value $V_{2c}$, which
depends on the value chosen for the other parameters $V_{4}(0)$ and
$V_{6}(0)$. The requirement is similar to the adjustment of the critical
temperature with a view to reach the critical point in actual systems.
The resulting value $V_{c}\left(\phi\right)$ is said to belong to
the critical subspace $\mathcal{S}_{c}$ of $\mathcal{S}$, which
forms the domain of attraction to the W-FP. $\mathcal{S}_{c}$ has
the same dimension as $\mathcal{S}$ minus one (the codimension of
$\mathcal{S}_{c}$ is equal to one) which corresponds to the fact
that the W-FP has only one direction of unstability, corresponding
to a direction locally orthogonal to $\mathcal{S}_{c}$ (we suppose
$h=0$).

It is thus possible to illustrate the RG trajectories in $\mathcal{S}_{c}$
which, starting from some arbitrary points of $\mathcal{S}$ reach
the W-FP when $s\rightarrow\infty$ (one initial parameter must be
finely adjusted). This is demonstrated in Figure 1. On this figure
one sees that, whatever their starting points, the critical trajectories
reach the W-FP asymptotically along a unique ideal trajectory which
links the G-FP to the W-FP. The closer the starting points are chosen
to the G-FP, the longer is the way along this ideal trajectory. This
ideal RG trajectory is called the {}``renormalized trajectory''
(RT) because the RG flow running on it corresponds, in the vicinity
of the G-FP, to the RG flow discovered in the perturbative theory
of renormalization. Since the RT is a manifold of dimension one, a
single parameter (the renormalized coupling {}``constant'' $g$)
is sufficient to characterize the RG flow running on it.

\begin{figure}
	\includegraphics[width=1\columnwidth]{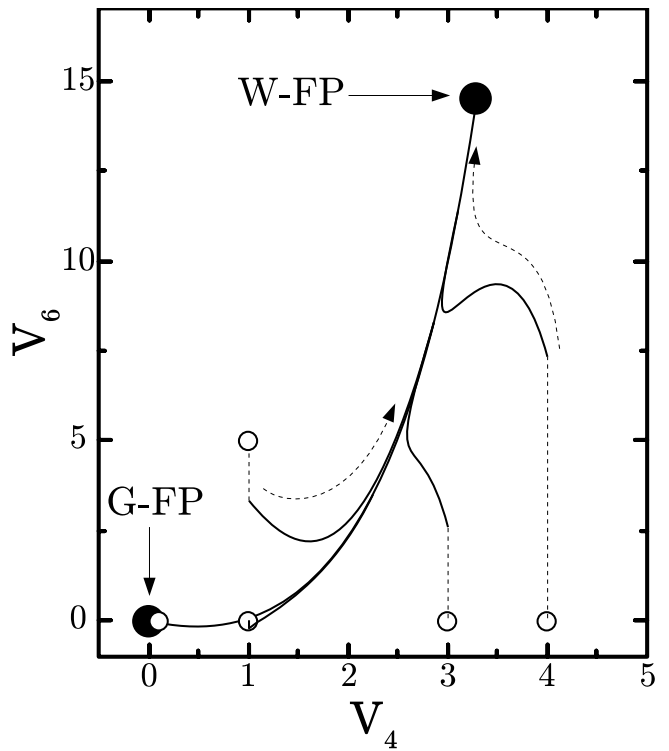}
	\caption{Ideal Renormalization Group trajectory in the binary diagram of coupling
parameters $V_{4}$ and $V_{6}$ {[}see text and Eqs. (\ref{dimensionless Wilson potential (19)})
and (\ref{WilPotInit (20)}){]}.
	\label{Figure 1}}
\end{figure}

Hence, in some sense (there is an arbitrariness in the definition
of the renormalized coupling) the renormalized (or running) coupling
$g$ of the perturbative framework implicitly {}``flows'' along
the RT. Actually, $g$ is a function of $s$, more precisely in the
circumstances of the massive field theory, it is a function of $\frac{\overline{m}}{\Lambda}$.
The difference with the complete Wilson's RG theory is that, in the
perturbative framework, one has implicitly assumed to be {}``exactly
on'' the RT, so that the distance to the W-FP is measured by the
values taken by a unique parameter $g\left(\frac{\overline{m}}{\Lambda}\right)-g^{\ast}$
($g^{\ast}$ being the value of $g$ at the W-FP). The consequence
is the presence of only one family of correction-to-scaling terms,
such as in the following expansion of the (dimensioless) correlation
length for example:

\begin{equation}
\ell^{*}\left(t^{\ast}\right)=\ell_{0}^{*\pm}\left|t^{\ast}\right|^{-\nu}\left[1+\sum_{k=1}^{\infty}b_{k}^{\pm}\left|t^{\ast}\right|^{k\Delta}\right]\label{FTSerieDeXI (21)}\end{equation}
 with $\Delta$ a specific exponent the value of which (close to $0.5$)
may be estimated by resumming series of the renormalized perturbation
theory of the $\Phi_{d}^{4}$ model. It is precisely expansions like
in Eq. (\ref{FTSerieDeXI (21)}) that have been summed into the crossover
functions considered above.

The other transients not accounted for by the $\Phi_{d}^{4}$ model
(by implicitly assuming that the only possible RG trajectory is the
RT) are responsible for other kinds of correction-to-scaling terms
characterized by a hierachical set of exponents $\Delta_{j}$ ($j=2,\cdots,\infty$)
such that: \[
\Delta<\Delta_{2}<\Delta_{3}<\cdots<\Delta_{\infty}\]

The fact that $\Delta$ (the exponent which controls the asymptotic
approach, along the RT, to the W-FP) is the smallest exponent is well
illustrated on Figure \ref{Figure 1} by the coincidence of all trajectories
with the RT before reaching the W-FP.

The few estimates of $\Delta_{2}$ indicate that its value in three
dimensions and for $n=1$ is of order $2\Delta$ so that, on a general
ground, the expansion (\ref{FTSerieDeXI (21)}) is a priori not valid
beyond the first correction term. This is why one often uses the following
two-term Wegner expansion to analyse the experimental data: \begin{equation}
\xi\left(\Delta\tau^{*}\right)=\xi_{0}^{\pm}\left|\Delta\tau^{*}\right|^{-\nu}\left[1+a_{\xi}^{\pm}\left|\Delta\tau^{*}\right|^{\Delta}\right]\label{2Terms (22)}\end{equation}
 with the exponents $\nu$ and $\Delta$ fixed to their FT values.
The domain of validity of such a two-term expansion is called the
preasymptotic domain (PAD) and noted $\left|t^{\ast}\right|\lesssim\mathcal{L}_{PAD}^{Ising}$.

For the same reason, the crossover functions calculated in FT were
not expected to generally reproduce the experimental data beyond the
Ising like PAD. However this objection does not account for the amplitudes
of the correction terms. It may well occur that the coefficients of
the corrections associated to the exponents $\Delta_{j}$ ($j=2,\cdots,\infty$)
are very small so that in effect only the $\Phi_{d}^{4}$-like corrections
have to be considered even on a range of $t^{\ast}$ for which several
(more than two) terms of the expansion of Eq. (\ref{FTSerieDeXI (21)})
are not negligible. It is the case of initial Hamiltonians the coordinates
of the critical point of which correspond to a point lying very close
to the RT. This particularity may even validate the entire crossover
functions of FT if the initial point lies close to the G-FP.

Of course, the RT does not represent the unique possibility of approaching
to the W-FP and the infinite dimension of $\mathcal{S}_{c}$ leaves
room for an infinity of possibilities. For example, there exists another
attractive trajectory which is characterized by an asymptotic approach
to the W-FP controlled by $\Delta_{2}$, and supplementary adjustment
of the initial Hamiltonian is needed to obtain this kind of approach.
For systems represented by such Hamiltonians, the crossover functions
of FT have no utility at all since none of the correction-to-scaling
terms is correct. Similar and more common are those systems which
correspond to an approach to the W-FP from the {}``opposite side''
compared to the RT. Their critical behavior are characterized by amplitudes
of the first correction-to-scaling (controlled by $\Delta$) with
a sign opposite to that generated by the approach along the RT \cite{Bagnuls1994}.

Up to now, it is not possible to say a priori which kind of approach
to the W-FP may correspond to an actual system which belongs to a
given universality class. A way to find out is to try a fit of test
functions to experimental data. From such tests, it seems that the
subclass of one-component fluids corresponds to Hamiltonian lying
very close to the RT \cite{Bagnuls2000}. This is why the crossover
functions of FT are good candidates to fit the scaling curves of that
subclass.

\section{The theoretical mean crossover functions}

In this section, we determine mean values of the crossover functions.
We also take the opportunity of simplifying the ad hoc functions by
limiting the products in Eq. (\ref{Function MR P (5)}) to only three
terms. Hence we have to determine the value of parameters like those
entering Eqs. (\ref{Function MR P (5)}) and (\ref{Deff exponent MR P (6)}),
but with $K=3$ for all quantities and so that the corresponding ad
hoc functions lie just in-between the max and min bounded functions
published in Ref. \cite{Bagnuls2002}. To accomplish this, the following
is taken into consideration.

\begin{figure}
	\includegraphics[width=1\columnwidth]{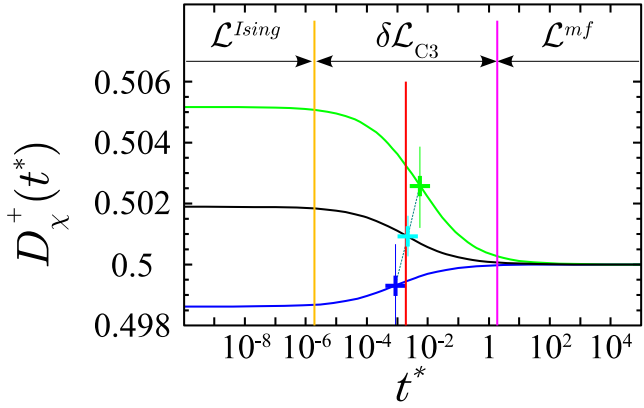}
	\caption{Typical crossover behaviors of the confluent crossover function $D_{\chi}^{+}\left(t^{*}\right)$
{[}Eq. (\ref{Deff exponent MR P (6)}){]}, for the bounded (max and
min) critical confluent exponents. as a function of the thermal field
(at zero ordering field) for the susceptibility in the homogenous
phase ($T>T_{c}$). The vertical lines are indicative of typical discrete
variations of the confluent function, showing the practical interest
of the condition $S_{\chi,2}^{+}\left(\Delta,\Delta_{mf}\right)\sqrt{t_{0}^{*}}=1$
(crosses), which corresponds to $D_{\chi}^{+}\left(t_{0}^{*}\right)=\Delta_{\frac{1}{2}}=\frac{\Delta+\Delta_{mf}}{2}$.
Using a single perfactor $\varpi<1$ which remains to be estimated
{[}see below Eq. (\ref{Ising PAD length vs S2 (40)}){]}, it is possible
to define the extensions $t^{*}<\mathcal{L}^{Ising}=\varpi\left(S_{\chi,2}^{+}\right)^{-2}$,
$t^{*}>\mathcal{L}^{mf}=\frac{1}{\varpi}\left(S_{\chi,2}^{+}\right)^{-2}$,
and $\mathcal{L}^{Ising}\lesssim t^{*}\in\left[\delta\mathcal{L}_{C3}\right]\lesssim\mathcal{L}^{mf}$,
of either, the Ising-like ($Ising$), or the mean-field-like ($mf$),
or the classical-to-critical crossover ($C3$) domains, respectively
(see text for details). \label{Figure 2}}
\end{figure}

\begin{enumerate}
\item The required precision in fitting each property requires the recourse
to phenomenological \emph{confluent} functions $D_{p}^{\pm}\left(t^{\ast}\right)$
{[}Eq.(\ref{Deff exponent MR P (6)}){]} which essentially account
for the \emph{crossover} between the values $\Delta\sim\frac{1}{2}$
of the Ising-like confluent exponent as $t^{\ast}\rightarrow0$, and
the exact value $\Delta_{mf}=\frac{1}{2}$ of the mean-field-like
{}``confluent'' exponent as $t^{\ast}\rightarrow\infty$. Such a
crossover introduces the following condition \begin{equation}
S_{P,1}^{\pm}=S_{P,2}^{\pm}\left(1-\Delta+\Delta_{mf}\right)\label{S1vS2 constraint MR P (23)}\end{equation}
 which confers to one parameter among $S_{P,1}^{\pm}$ and $S_{P,2}^{\pm}$,
a noticeable difference from the other constants appearing in Eqs.
(\ref{Function MR P (5)}) and (\ref{Deff exponent MR P (6)}). 
\item In \cite{Bagnuls2002}, the ad hoc functions for the susceptibility
in the homogeneous phase are already defined with a three-term product.
In that case, the behaviors of $D_{\chi}^{+}\left(t^{\ast}\right)$
as a function of $t^{\ast}$ as illustrated in Figure \ref{Figure 2},
clearly show that the condition \[
S_{\chi,2}^{\pm}\sqrt{t_{0}^{\ast}}=1\]

can be used as an indicative sensor of the classical-to-critical crossover
(C3) domain ($t_{0}^{\ast}$ stands for the values of $t^{*}$ where
the ad hoc confluent functions take on precisely the mean value $\Delta_{\frac{1}{2}}=\frac{\Delta+\Delta_{mf}}{2}$).
The extension $\delta\mathcal{L}_{C3}$ of the associated {}``intermediate''
$t^{\ast}$-range can be measured by a characteristic amplitude $\varpi<1$
{[}which remains to be defined, see below Eq. (\ref{tstardelta confluent condition vs S2 (39)}){]},
such that to $\varpi\left(S_{\chi,2}^{\pm}\right)^{-2}\lesssim t^{\ast}\in\left[\delta\mathcal{L}_{C3}\right]\lesssim\frac{1}{\varpi}\left(S_{\chi,2}^{\pm}\right)^{-2}$,
separating thus unambiguously an Ising-like asymptotic domain of extension
$t^{\ast}<\mathcal{L}^{Ising}=\varpi\left(S_{\chi,2}^{\pm}\right)^{-2}$
and a mean-field like asymptotic domain of extension $t^{\ast}>\mathcal{L}^{mf}=\frac{1}{\varpi}\left(S_{\chi,2}^{\pm}\right)^{-2}$
(see Figure \ref{Figure 2}).

\item The leading amplitudes $\mathbb{Z}_{P}^{\pm}$, associated with their
respective universal exponents $e_{P}$ satisfying scaling laws, are
unambiguously related by universal amplitude combinations between
them. 
\item A part of $\left\{ X_{i},Y_{i}\right\} _{P}^{\pm}$'s accounts for
the universal features associated to the critical confluent corrections
to scaling, explicitly characterized by the single universal exponent
$\Delta$ and one first confluent amplitude (see below) for $t^{\ast}\rightarrow0$. 
\item Another part accounts for the asymptotic mean-field behavior for $t^{\ast}\rightarrow\infty$
{[}implicitly characterized by the two (leading and confluent) exponents
$e_{P,mf}$ and $\Delta_{mf}$ and their associated amplitudes (see
below){]}. 
\item The remaining part accounts for the expected classical-to-critical
crossover (C3) in the intermediate $t^{\ast}$-range (which also remains
to be defined, see point 2 just above). 
\end{enumerate}

These remarks provide constraints on the parameter entering each specific
ad hoc functions. Let us consider those constraints explicitly.

\subsection{3D Ising-like PAD description}

Since the error estimates are the largest in the vicinity of the W-FP,
it is in the Ising-like PAD limit that the determination of the mean
value has important consequences.

Let us define a set of four constraints to control the \emph{mean}
Ising-like PAD description within the two limiting Ising-like PAD
descriptions given by the bounded functions $F_{P,max}$ and $F_{P,min}$,
respectively.

Starting with Eq. (\ref{Function MR P (5)}), we first consider the
two-term (Wegner) expansion of $F_{P}\left(t^{\ast}\right)$: \begin{equation}
F_{P,Ising}^{PAD}\left(t^{\ast}\right)=\mathbb{Z}_{P}^{\pm}\left(t^{\ast}\right)^{-e_{P}}\left[1+\mathbb{Z}_{P}^{1,\pm}\left(t^{\ast}\right)^{\Delta}\right]\label{FPADIsing MR P (24)}\end{equation}
 where the function $F_{P,Ising}^{PAD}\left(t^{\ast}\right)$ is valid
within some Ising-like PAD extension $t^{\ast}\leq\mathcal{L}_{PAD}^{Ising}$.
To estimate mean exponents and mean amplitudes of Eq. (\ref{FPADIsing MR P (24)})
from max and min ones, we impose the following four obvious conditions
\begin{equation}
e_{P}=\frac{e_{P,max}+e_{P,min}}{2}\label{ePIsing MR (25)}\end{equation}
\begin{equation}
\mathbb{Z}_{P}^{\pm}=\sqrt{\mathbb{Z}_{P,max}^{\pm}\mathbb{Z}_{P,min}^{\pm}}\label{ZPIsing MR (26)}\end{equation}
\begin{equation}
\Delta=\frac{\Delta_{max}+\Delta_{min}}{2}\label{deltaIsing MR (27)}\end{equation}
\begin{equation}
\begin{array}{cl}
\mathbb{Z}_{P}^{1,\pm} & =\sum_{i=1}^{3}X_{P,i}^{\pm}Y_{P,i}^{\pm}\\
 & =\sum_{i=1}^{K}\frac{X_{P,max,i}^{\pm}Y_{P,max,i}^{\pm}+X_{P,min,i}^{\pm}Y_{P,min,i}^{\pm}}{2}\end{array}\label{Z1PIsing MR (28)}\end{equation}

The leading term of our mean function is then identical to the leading
term of the mixing function \begin{equation}
\begin{array}{cl}
F_{P,mix}^{E}\left[t^{\ast},D_{P,mix}^{\pm}\left(t^{\ast}\right)\right] & =\left\{ F_{P,max}\left[t^{\ast},D_{P,max}^{\pm}\left(t^{\ast}\right)\right]\right\} ^{E}\\
 & \times\left\{ F_{P,min}\left[t^{\ast},D_{P,max}^{\pm}\left(t^{\ast}\right)\right]\right\} ^{1-E}\end{array}\label{FEmix MR P (29)}\end{equation}
 proposed in \cite{Bagnuls2002}, with $E=\frac{1}{2}$. The three
first Eqs. (\ref{ePIsing MR (25)}), (\ref{ZPIsing MR (26)}), and
(\ref{deltaIsing MR (27)}), provide unequivocal determination of
the mean values of the three parameters $e_{P}$, $\mathbb{Z}_{P}^{\pm}$,
and $\Delta$, respectively. The Eq. (\ref{Z1PIsing MR (28)}) involves
exclusively the $\left\{ X_{i},Y_{i}\right\} _{P}^{\pm}$'s.

\subsection{Mean-field-like PAD description}

To recover the (asymptotic) mean-field-like behavior of $F_{P}\left[t^{\ast},D_{P}^{\pm}\left(t^{\ast}\right)\right]$
in the limit $t^{\ast}\rightarrow\infty$, the following relations
are required

\begin{equation}
	e_{P,mf}=e_{P}-\frac{1}{2}\sum_{i=1}^{3}Y_{P,i}^{\pm}
	\label{emf-e constraint MR P (30)}
\end{equation}
\begin{equation}
	\mathbb{Z}_{P,mf}^{\pm} = \mathbb{Z}_{P}^{\pm} \prod_{i=1}^{3} \left( X_{P,i}^{\pm} \right)^{Y_{P,i}^{\pm}}
	\label{Zmf amplitude MR P (31)}
\end{equation}
 where $e_{P,mf}$ and $\mathbb{Z}_{P,mf}^{\pm}$ are the mean-field
(classical) values of the asymptotic exponent and amplitude, respectively.
Such a mean-field situation as $t^{\ast}\rightarrow\infty$, can be
easily characterized in an equivalent manner to the above Ising situation
as $t^{\ast}\rightarrow0$.

Therefore, the following restricted two-term expansions \begin{equation}
F_{P,mf}^{PAD}\left(t^{\ast}\right)=\mathbb{Z}_{P,mf}^{\pm}\left(t^{\ast}\right)^{-e_{mf}}\left[1+Z_{P,mf}^{1,\pm}\left(t^{\ast}\right)^{-\Delta_{mf}}\right]\label{FPADmf MR P (32)}\end{equation}
 can be easily determined for $t^{\ast}\rightarrow\infty$, which
are valid close to the G-FP, within the mean-field-like PAD extension
$t^{\ast}\geq\mathcal{L}_{PAD}^{mf}$. In Eq. (\ref{FPADmf MR P (32)}),
$\Delta_{mf}$ is the mean field exponent of the first order term
of {}``classical'' corrections, and $\mathbb{Z}_{P,mf}^{1,\pm}$
is the associated amplitude defined by the following equation

\begin{equation}
\mathbb{Z}_{P,mf}^{1,\pm}=\sum_{i=1}^{3}\frac{Y_{P,i}^{\pm}}{X_{P,i}^{\pm}}\label{Z1mf amplitude MR P (33)}\end{equation}

As in the case of the Ising-like PAD description, we impose four conditions
on the ad hoc mean function to describe the mean-field PAD using the
original pairs $\left\{ X_{i},Y_{i}\right\} _{P}^{\pm}$'s of the
bounded functions. These four conditions read as follows \begin{equation}
\begin{array}{cl}
e_{P,mf} & =e_{P}-\frac{1}{2}\sum_{i=1}^{3}Y_{P,i}^{\pm}\\
 & =\frac{1}{2}\left[e_{P,max}+e_{P,min}\right.\\
 & \left.-\frac{1}{2}\sum_{i=1}^{K}\left(Y_{P,max,i}^{\pm}+Y_{P,min,i}^{\pm}\right)\right]\end{array}\label{ePmf MR (34)}\end{equation}
\begin{equation}
\begin{array}{cl}
\mathbb{Z}_{P,mf}^{\pm} & =\mathbb{Z}_{P}^{\pm}\prod_{i=1}^{3}\left(X_{P,i}^{\pm}\right)^{Y_{P,i}^{\pm}}\\
 & =\sqrt{\mathbb{Z}_{P,max}^{\pm}\mathbb{Z}_{P,min}^{\pm}}\\
 & \times\prod_{i=1}^{K}\sqrt{\left(X_{P,max,i}^{\pm}\right)^{Y_{P,max,i}^{\pm}}\left(X_{P,min,i}^{\pm}\right)^{Y_{P,min,i}^{\pm}}}\end{array}\label{Zpmf MR (35)}\end{equation}
\begin{equation}
\Delta_{P,mf}=\frac{1}{2}\label{deltamf MR (36)}\end{equation}

\begin{equation}
\begin{array}{cl}
\mathbb{Z}_{P,mf}^{1,\pm} & =\sum_{i=1}^{3}\frac{Y_{P,i}^{\pm}}{X_{P,i}^{\pm}}\\
 & =\frac{\lambda_{P}}{2}\sum_{i=1}^{K}\left(\frac{Y_{P,max,i}^{\pm}}{X_{P,max,i}^{\pm}}+\frac{Y_{P,min,i}^{\pm}}{X_{P,min,i}^{\pm}}\right)\end{array}\label{Z1Pmf MR (37)}\end{equation}

Eq. (\ref{deltamf MR (36)}) fixes the mean value of the mean-field-like
{}``confluent'' exponent, unequivocally. The three Eqs. (\ref{ePmf MR (34)}),
(\ref{Zpmf MR (35)}), and (\ref{Z1Pmf MR (37)}), added to Eq. (\ref{Z1PIsing MR (28)}),
impose the mean values of two pairs among the $3$ pairs $\left\{ X_{i},Y_{i}\right\} _{P}^{\pm}$'s.

\subsection{Number and nature of the parameters}

The above analysis demonstrates that the theoretical mean function
$F_{P}\left[t^{\ast},D_{P}^{\pm}\left(t^{\ast}\right)\right]$ must
satisfy $9$ constraints {[}Eqs. (\ref{S1vS2 constraint MR P (23)}),
(\ref{ePIsing MR (25)}) to (\ref{Z1PIsing MR (28)}), and (\ref{ePmf MR (34)})
to (\ref{Z1Pmf MR (37)}){]} to reproduce the two asymptotic and pre-asymptotic
branches of a complete crossover effect. Therefore, any $F_{P}\left[t^{\ast},D_{P}^{\pm}\left(t^{\ast}\right)\right]$
must contain at least $9$ parameters.

Eight of them can be readily unequivocally determined: $e_{P}$ from
Eq. (\ref{ePIsing MR (25)}), $\mathbb{Z}_{P}^{\pm}$ from Eq. (\ref{ZPIsing MR (26)}),
$\Delta$ from Eq. (\ref{deltaIsing MR (27)}), $\Delta_{mf}$ from
Eq. (\ref{deltamf MR (36)}) and two $\left\{ X_{i},Y_{i}\right\} _{P}^{\pm}$
pairs from the four Eqs. (\ref{Z1PIsing MR (28)}), (\ref{ePmf MR (34)}),
(\ref{Zpmf MR (35)}), and (\ref{Z1Pmf MR (37)}). Using $S_{P,2}^{\pm}\left(\Delta,\Delta_{mf}\right)$
as an entry data in Eq. (\ref{Deff exponent MR P (6)}), leads to
the supplementary determination of $S_{P,1}^{\pm}\left(\Delta,\Delta_{mf}\right)$
from Eq. (\ref{S1vS2 constraint MR P (23)}). Subsequently, the parameters
attached to the intermediate part of the crossover need to be determined.

The phenomenological forms of Eqs. (\ref{Function MR P (5)}), (\ref{Deff exponent MR P (6)})
and (\ref{S1vS2 constraint MR P (23)}), with $K=3$, introduce $12$
parameters {[}the case with $K=2$ ($10$ parameters) is of limited
interest due to the error-bar propagation which cannot be accounted
for via only two $\left\{ X_{i},Y_{i}\right\} _{P}^{\pm}$ pairs{]}.
Therefore, for any fitting procedure which uses $S_{P,2}^{\pm}\left(\Delta,\Delta_{mf}\right)$
as an entry (free) parameter, only one $\left\{ X_{i},Y_{i}\right\} _{P}^{\pm}$
pair, remains truly free in the $\left(K=3\right)$ product terms
of the MR crossover function.

The triad $\left\{ S_{2},X_{i},Y_{i}\right\} _{P}^{\pm}$ of the calculated
parameters is composed of one {}``crossover sensor'' $S_{P,2}^{\pm}$,
characteristic of the $t^{\ast}$-\emph{location} of the classical-to-critical
confluent crossover (see Figure \ref{Figure 2}), and one {}``amplitude-exponent
pair'' $\left\{ X_{i},Y_{i}\right\} _{P}^{\pm}$, proper to the \emph{shape}
of the strict crossover part between the two asymptotic PAD behaviors
(we will illustrate this latter point below using Figure \ref{Figure 4}
in § 3.4).

Let us look for the possible existence of a unique value $S_{2}=cte$
{[}or $S_{1}=cte$ by virtue of Eq. (\ref{S1vS2 constraint MR P (23)}){]},
whatever the selected property and the considered (homogeneous or
non-homogeneous) phase of the system.

\begin{figure}
\includegraphics[%
  width=1\columnwidth]{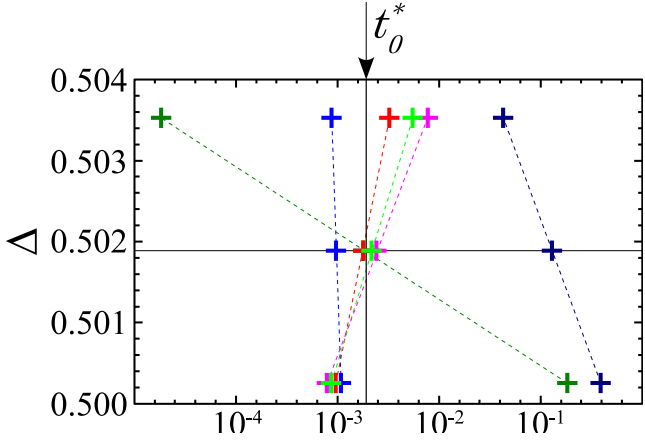}

\caption{Characteristic values of the crossover condition $S_{P,2}^{\pm}\left(\Delta,\Delta_{mf}\right)\sqrt{t^{*}}=1$,
for all the {}``min'' (upper crosses at same $\Delta_{max}$ value)
and {}``max'' (lower crosses at same $\Delta_{min}$ value) crossover
functions estimated in Ref. \cite{Bagnuls2002} (lin-log $\Delta$;$t^{*}$
diagram). The $P$ (property) color indexations are: green $\chi^{*}\left(+\right)$,
red $\ell^{*}\left(+\right)$, blue $C^{*}\left(+\right)$, pink $m^{*}\left(-\right)$,
dark green $\chi^{*}\left(-\right)$, and dark blue $C^{*}\left(-\right)$.
The corresponding values for the mixing crossover functions with $E=\frac{1}{2}$
{[}see Eq. (\ref{FEmix MR P (29)}){]}, are indicated by the intermediate
crosses (at same $\Delta_{mean}=\frac{\Delta_{max}+\Delta_{min}}{2}$
value). The linear segments show the agreement with the associated
geometrical mean values $\sqrt{S_{P,max,2}^{\pm}S_{P,min,2}^{\pm}t^{*}}=1$
of the min and max crossover conditions. The vertical line illustrates
the {}``universal'' crossover condition $S_{2}\sqrt{t_{0}^{*}}=1$
{[}where $D\left(\frac{1}{\left[S_{2}\left(\Delta_{mean},\Delta_{mf}\right)\right]^{2}}\right)=\frac{\Delta_{mean}+\Delta_{mf}}{2}$
from Eq. (\ref{Deff exponent MR P (6)}){]}, which is selected in
the present study, whatever the property $\left(P\right)$ or the
phase domain $\left(\pm\right)$. \label{Figure 3}}
\end{figure}

The $S_{P,i}^{\pm}\left(\Delta,\Delta_{mf}\right)$ parameters are
only related to the two universal confluent exponents. However, the
numerical values of $\Delta$ appear also conditioned by the error-bar
propagation of asymptotic uncertainties provided by the theoretical
estimations of the leading universal exponents close to the non-Gaussian
fixed point. In Figure \ref{Figure 3} we have reported by crosses
at constant $\frac{\Delta+\Delta_{mf}}{2}$ as a function of $t^{*}$,
all the {[}max, min, and mixing{]} conditions $S_{P,2}^{\pm}\sqrt{t^{*}}=1$,
whatever $P$ and $\pm$ states (extending then the previous results
reported in Figure \ref{Figure 2} for the homogeneous susceptibility
case).

Since the dispersion of the mixing conditions $\sqrt{S_{P,max,2}^{\pm}S_{P,min,2}^{\pm}t^{\ast}}=1$
is significatively lowered, Figure 3 supports the possibility of choosing
a unique value for $S_{2}$.

Moreover, anticipating the $S_{2}$ estimation given in the following
subsection, the resulting \emph{mean} value $S_{2}=22.9007$, common
to all the properties and all the states, provides a condition $S_{2}\sqrt{t_{0}^{\ast}}=1$
in close agreement with the previous mixed conditions, as shown by
the corresponding vertical line in Figure \ref{Figure 3}. The departure
of the non-homogenous heat capacity case is probably due to the cumulative
effect of the error bar propagation of the uncertainties on the exponent
and on the critical background amplitude estimations, related here
to the \emph{non-zero} value of the regular background term (equal
to $3$) below $T_{c}$. We can then note that the above well-controlled
origin of this {}``universal'' model-parameter $S_{2}$, partly
compensates for the arbitrariness of its numerical value.

\begin{table}
\begin{tabular}{|c|c|c|c|c|}
\hline 
$0$&
 $P$&
 $1/\ell^{*}$&
 $1/\chi^{*}$&
 $C^{*}$\tabularnewline
\hline
$1$&
 $e_{P}^{+}$&
 $-0.6303875$&
 $-1.2395935$&
 $0.1088375$\tabularnewline
\hline
$2$&
 $\mathbb{Z}_{P}^{+}$&
 $2.121008$&
 $3.709601$&
 $1.719788$\tabularnewline
\hline
$3$&
 $\Delta$&
 $0.50189$&
&
\tabularnewline
\hline
$4$&
 $\mathbb{Z}_{P}^{1,+}$&
 $-5.81623$&
 $-8.56347$&
 $8.06569$\tabularnewline
\hline
$5$&
 $S_{1}$&
 $22.8573$&
&
\tabularnewline
\hline
$6$&
 $S_{2}$&
 $22.9007$&
&
\tabularnewline
\hline
$7$&
 $X_{P,1}$&
 $40.0606$&
 $29.1778$&
 $36.6874$\tabularnewline
\hline
$8$&
 $Y_{P,1}$&
 $-0.098968$&
 $-0.178403$&
 $0.220033$\tabularnewline
\hline
$9$&
 $X_{P,2}$&
 $11.93211$&
 $11.7625$&
 $3.23787$\tabularnewline
\hline
$10$&
 $Y_{P,2}$&
 $-0.153912$&
 $-0.282241$&
 $-0.000133095$\tabularnewline
\hline
$11$&
 $X_{P,3}$&
 $1.902735$&
 $2.05948$&
 $2.84102$\tabularnewline
\hline
$12$&
 $Y_{P,3}$&
 $-0.00789505$&
 $-0.0185424$&
 $-0.00222489$\tabularnewline
\hline
$13$&
 $X_{C}$&
&
&
 $-3.79829$\tabularnewline
\hline
$14$&
 $e_{P,mf}$&
 $-0.5$&
 $-1$&
 $0$\tabularnewline
\hline
$15$&
 $\mathbb{Z}_{P,mf}^{+}$&
 $1$&
 $1$&
 $3.79004$\tabularnewline
\hline
$16$&
 $\Delta_{mf}$&
 $-0.5$&
&
\tabularnewline
\hline
$17$&
 $\mathbb{Z}_{P,mf}^{1,+}$&
 $-0.0195196$&
 $-0.0391128$&
 $0.00517327$ \tabularnewline
\hline
\end{tabular}

\caption{Numerical values of the parameters of the mean crossover functions
$F_{P}\left[t^{*},D\left(t^{*}\right)\right]$ {[}see Eqs. (\ref{Function MR P (5)})
and (\ref{Deff exponent MR P (6)}), with $K=3$ and $S_{P,2}^{\pm}=S_{2}${]},
corresponding to the dimensionless correlation length $\ell^{*}$,
susceptibility $\chi^{*}$, and specific heat $C^{*}$, in the homogeneous
phase ($T>T_{c}$). The lines $1$ to $4$ correspond to the characteristic
parameters of the Ising-like PAD description {[}see Eqs. (\ref{FPADIsing MR P (24)})
to (\ref{Z1PIsing MR (28)}){]}, while the lines $14$ to $17$ correspond
to the characteristic parameters of the mean field-like PAD description
{[}see Eqs. (\ref{FPADmf MR P (32)}) to (\ref{Z1Pmf MR (37)}{]}.
The universal values of the interrelated crossover sensors $S_{1}$
and $S_{2}$ {[}see Eqs. (\ref{S1vS2 constraint MR P (23)}){]}, are
given in the lines $5$ and $6$. The values of the three amplitude-exponent
pairs $\left\{ X_{i},Y_{i}\right\} _{P}^{+}$ given in lines $7$
to $12$, have been determined by a carefull adjustement to the theoretical
mixing functions of Eq. (\ref{FEmix MR P (29)}) (with $E=\frac{1}{2}$)
proposed in Ref. \cite{Bagnuls2002}, using specific constraints of
Equations (\ref{ePIsing MR (25)}) to (\ref{Z1PIsing MR (28)}) and
(\ref{ePmf MR (34)}) to (\ref{Z1Pmf MR (37)}). The value of the
critical background $X_{C}$ of the specific heat given in line $13$
is the mean value of the min and max values of parameters $X_{6}$
of Ref. \cite{Bagnuls2002}. \label{Table I}}
\end{table}

\begin{table}
\begin{tabular}{|c|c|c|c|c|c|}
\hline 
$0$&
 $P$&
 $m^{*}$ (MR7)&
 $m^{*}$ (MR67)&
 $1/\chi^{*}$&
 $C^{*}$\tabularnewline
\hline
$1$&
 $e_{P}^{-}$&
 $0.3257845$&
 $0.3257845$&
 $-1.2395935$&
 $0.1088375$\tabularnewline
\hline
$2$&
 $\mathbb{Z}_{P}^{-}$&
 $0.937528$&
 $0.937528$&
 $17.762821$&
 $3.203771$\tabularnewline
\hline
$3$&
 $\Delta$&
 $0.50189$&
 $0.50189$&
&
\tabularnewline
\hline
$4$&
 $\mathbb{Z}_{P}^{1,-}$&
 $3.42538$&
 $7.70712$&
 $-40.4666$&
 $6.69984$\tabularnewline
\hline
$5$&
 $S_{1}$&
 $22.8573$&
 $22.8573$&
&
\tabularnewline
\hline
$6$&
 $S_{2}$&
 $22.9007$&
 $22.9007$&
&
\tabularnewline
\hline
$7$&
 $X_{P,1}$&
 $219.597$&
 $124.274$&
 $470.671$&
 $116.515$\tabularnewline
\hline
$8$&
 $Y_{P,1}$&
 $-0.0284374$&
 $0.0206744$&
 $-0.0332665$&
 $-0.105953$\tabularnewline
\hline
$9$&
 $X_{P,2}$&
 $38.3772$&
 $20.9461$&
 $43.6468$&
 $72.9532$\tabularnewline
\hline
$10$&
 $Y_{P,2}$&
 $0.224464$&
 $0.20453$&
 $-0.585082$&
 $0.247902$\tabularnewline
\hline
$11$&
 $X_{P,3}$&
 $6.92804$&
 $6.92808$&
 $5.2264$&
 $12.6739$\tabularnewline
\hline
$12$&
 $Y_{P,3}$&
 $0.152404$&
 $0.123226$&
 $0.139161$&
 $0.0757262$\tabularnewline
\hline
$13$&
 $X_{C}$&
&
&
&
 $-3.79349$\tabularnewline
\hline
$14$&
 $e_{P,mf}$&
 $0.5$&
 $0.5$&
 $-1$&
 $0$\tabularnewline
\hline
$15$&
 $\mathbb{Z}_{P,mf}^{-}$&
 $\sqrt{6}$&
 $\sqrt{6}$&
 $2$&
 $6.79349$\tabularnewline
\hline
$16$&
 $\Delta_{mf}$&
 $-0.5$&
 $-0.5$&
&
\tabularnewline
\hline
$17$&
 $\mathbb{Z}_{P,mf}^{1,-}$&
 $0.0277176$&
 $0.0277175$&
 $0.0131516$&
 $0.00846373$ \tabularnewline
\hline
\end{tabular}

\caption{Same as Table I for the dimensionless order parameter $m^{\ast}$,
susceptibility $\chi^{\ast}$, and specific heat $C^{\ast}$, in the
non-homogeneous phase ($T<T_{c}$). The distinction between the MR7
and MR 67 results accounts for difference in the numerical values
of the first confluent amplitude $\mathbb{Z}_{M}^{1}$ of Eq. (\ref{Z1PIsing MR (28)}),
which provides the (central) universal values $\frac{\mathbb{Z}_{M}^{1}}{\mathbb{Z}_{\chi}^{1,+}}\left(MR7\right)=0.45$
and $\frac{\mathbb{Z}_{M}^{1}}{\mathbb{Z}_{\chi}^{1,+}}\left(MR67\right)\equiv\frac{\mathbb{Z}_{M}^{1}}{\mathbb{Z}_{\chi}^{1,+}}\left(MR6\right)=0.9$
(see Appendix B for details). \label{Table II}}
\end{table}

\subsection{Unique form of the mean \emph{confluent} function}

For each property, the three associated functions,\[
F_{P,th}\left[t^{\ast},D_{P}^{\pm}\left(t^{\ast}\right)\right]=\left\{ \begin{array}{c}
F_{P,max}\left[t^{\ast},D_{P,max}^{\pm}\left(t^{\ast}\right)\right]\\
F_{P,min}\left[t^{\ast},D_{P,max}^{\pm}\left(t^{\ast}\right)\right]\\
F_{P,mix}^{E=\frac{1}{2}}\left[t^{\ast},D_{P,mix}^{\pm}\left(t^{\ast}\right)\right]\end{array}\right\} ,\]
 proposed in reference \cite{Bagnuls2002} can be considered. The
three corresponding residual functions, $r_{P}\left(t^{\ast}\right)$,
expressed in \%, are calculated from reference to our mean function
noted $F_{P}\left[t^{\ast},D\left(t^{\ast}\right)\right]$, such as
$r_{P}\left(t^{\ast}\right)=100\times\left(\frac{F_{P,th}\left[t^{\ast},D_{P}^{\pm}\left(t^{\ast}\right)\right]}{F_{P}\left[t^{\ast},D\left(t^{\ast}\right)\right]}-1\right)$.
The results reported hereafter are obtained from the minimization
method of the residuals related only to the mixing function with $E=\frac{1}{2}$
{[}see Eq. \ref{FEmix MR P (29)}{]}.

In a first step, starting with the susceptibility in the homogeneous
phase as a basic property (since $K=3$ already), we have validated
the derivation of several mean functions using several entry triads
made of $\sqrt{S_{\chi,max,2}^{+}S_{\chi,min,2}^{+}}$ and one among
the three pairs $\left\{ X_{i,max},Y_{i,max}\right\} _{\chi}^{+}$,
$\left\{ X_{i,min},Y_{i,min}\right\} _{\chi}^{+}$, or $\left\{ \sqrt{X_{i,max}X_{i,min}},\frac{Y_{i,max}+Y_{i,min}}{2}\right\} _{\chi}^{\pm}$
for each $i$ value between $1$ to $K$. The two remaining pairs
$\left\{ X_{j\neq i},Y_{j\neq i}\right\} _{\chi}^{+}$s were then
calculated from Eqs.( \ref{Z1PIsing MR (28)}), (\ref{ePmf MR (34)}),
(\ref{Zpmf MR (35)}) and (\ref{Z1Pmf MR (37)}). This step was then
repeated for all the properties with two main results:

\begin{enumerate}
\item at least one solution with $K=3$ exists for all the properties and all the states; 
\item the residuals are minimum for the triad $\left\{ \sqrt{S_{max,2}S_{min,2}},\sqrt{X_{i,max}X_{i,min}},\frac{Y_{i,max}+Y_{i,min}}{2}\right\} _{P}^{\pm}$. 
\end{enumerate}
The following complementary observations were also made:

\begin{enumerate}
\item The significative values of residuals are in a $t^{\ast}$-range where
$S_{P,2}^{\pm}\sqrt{t^{\ast}}\simeq1$. 
\item The $t^{\ast}$-range where our mean functions compare to the max
(or min) functions, corresponds to $t^{\ast}>\frac{0.1}{S_{P,2}^{\pm}}$
(in this domain, the residuals are of the same order of magnitude
than those obtained using the ($E=\frac{1}{2}$) mixing functions
and remains $\lesssim10^{-3}$). 
\item Each Ising-like PAD description by equation (\ref{FPADIsing MR P (24)})
agrees with the complete crossover function within an error-bar lower
than $0.01$\% for $t^{\ast}\leq\mathcal{L}_{PAD}^{Ising}$, where
$\mathcal{L}_{PAD}^{Ising}$ is then defined by the value\begin{equation}
\left(S_{P,2}^{\pm}\right)^{2}\mathcal{L}_{PAD}^{Ising}\simeq10^{-3}\label{SP2-Ising PAD length combination (38)}\end{equation}

\end{enumerate}
In a second step, for the inverse susceptibility and the inverse correlation
length in the homogeneous state, we have minimized the residuals by
successive small variations around the above mixing values of each
constitutive parameter of the entry triad. We have then selected the
\emph{best} mean functions for each property, using the mean value
$S_{2}=22.9007$ of the two optimized values $S_{\chi,2}=22.9321$
and $S_{\xi,2}=22.8693$, associated to $\chi$ and $\xi$, respectively.
The final step was to minimize all the residuals with $S_{2}=22.9007$
fixed, whatever the property or the domain. As a main consequence,
the universal confluent crossover condition $S_{2}\sqrt{t^{*}}=1$
is such that $t^{*}=t_{0}^{*}$, with\begin{equation}
t_{0}^{*}=\frac{1}{\left(S_{2}\right)^{2}}\cong2.\,10^{-3}\label{tstardelta confluent condition vs S2 (39)}\end{equation}
 Accordingly, Eq. (\ref{SP2-Ising PAD length combination (38)}) has
a universal form whatever the property, leading to well-controlled
PAD extension $t^{\ast}\leq\mathcal{L}_{PAD}^{Ising}$, with\begin{equation}
\mathcal{L}_{PAD}^{Ising}\simeq\frac{10^{-3}}{\left(S_{2}\right)^{2}}=\varpi\times t_{0}^{*}\cong2.\,10^{-6}\label{Ising PAD length vs S2 (40)}\end{equation}
 where $\varpi\cong10^{-3}$ is then a convenient parameter to account
for the intermediate crossover range between the two (Ising like and
mean field like) PADs (see Figure \ref{Figure 2}, for the susceptibility
case as a typical example ).

The corresponding parameters of the theoretical mean functions are
given in Table \ref{Table I} (homogeneous state, $\Delta\tau^{\ast}>0$)
and Table \ref{Table II} (non-homogeneous state, $\Delta\tau^{\ast}<0$).
The mean crossover functions for the susceptibility, the correlation
length and specific heat in the homogeneous domain are within the
above theoretical level of precision in the complete $t^{\ast}$-range.
For the order parameter, susceptibility and specific heat in the non-homogeneous
domain we get a sufficient accuracy ($\pm0.1$\%) to have agreement
with the theoretical crossover shape in the intermediate $t^{\ast}$-range.
The MR67 order parameter case in the non-homogeneous domain is the
object of a specific analysis (see Appendix B) related to the application
restricted to the one component fluid subclass. Correspondingly, for
a more detailed analysis on the level of precision, we have also reported
in this Appendix B the residuals for all the properties (see Figures
\ref{Figure 5} to \ref{Figure 10}).

To evaluate the influence of the constant universal value $S_{2}=cte=22.9007$,
we have made a comparison between all the $t_{e_{P}}^{\ast}$-values
where the following local mean value\begin{equation}
e_{P,eff}=e_{P,\frac{1}{2}}=\frac{e_{P}+e_{mf}}{2}\label{eP 1d2 (25)}\end{equation}
 of the effective exponent $e_{P,eff}$ occurs for any max, min, and
mixing crossover function {[}the effective exponent is given by the
equation $e_{P,eff}=-\frac{\partial\ln\left[F_{P}\left[t^{\ast},D_{P}^{\pm}\left(t^{\ast}\right)\right]\right]}{\partial\ln t^{*}}$
\cite{Kouvel1964}{]}.

\begin{figure}[ht]
	\includegraphics[width=1\columnwidth]{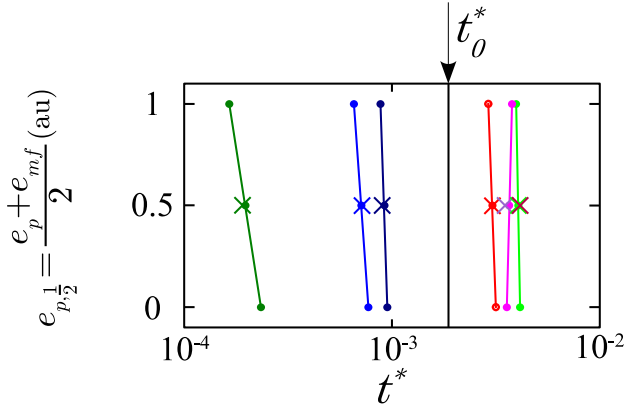}
	\caption{$t_{e_{P}}^{*}$-position (semi-log scale) of the effective mean
value $e_{P,\frac{1}{2}}=\frac{e_{P}+e_{mf}}{2}$, (in arbitrary units),
between critical and classical asymptotic exponents. Upper, lower,
and median circles are for $e_{P}$ max, min, and mixing values of
reference \cite{Bagnuls2002}, while median crosses are for mean values
of the present mean crossover functions. See the legend of Figure
3 for the $P$ color indexation. The vertical (black) line indicates
the $t_{0}^{*}$-position for the universal confluent condition $S_{2}\sqrt{t_{0}^{*}}=1$.
Note the significative differences in $t_{e_{P}}^{*}$-positions for
max, min, and mean values of the effective leading exponents for each
property, compared to the {}``universal{}`` effective confluent
exponent case reported in Figure \ref{Figure 3}.
	\label{Figure 4}}
\end{figure}

\begin{figure*}
	\includegraphics[width=2\columnwidth]{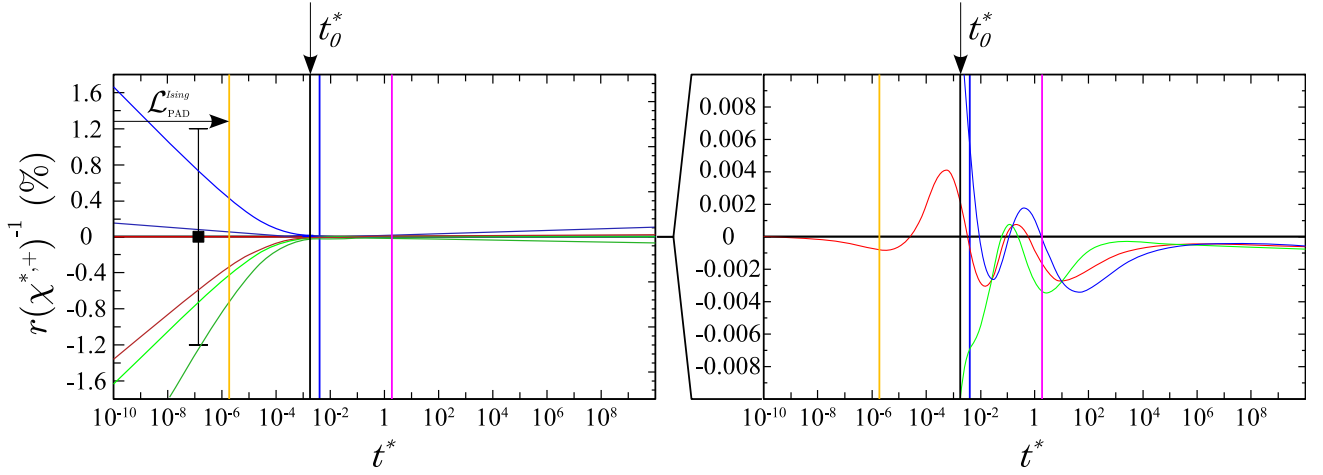}
	\caption{Residuals (\%) for the inverse susceptibility in the homogeneous
phase. The right part compares relative residuals to the theoretical
fitting precision of $0.01$\%. Color indexation: green (dark green)
: MR7 (MR6) min crossover function; blue (dark blue): MR7 (MR6) max
crossover function; red (dark red): MR7 (MR6) mixing crossover function
with $E=\frac{1}{2}$; Residuals for MR6 functions are not visible
in the magnified rigth scale, except in the C3 range. Four characteristics
$t^{*}$-positions are indicated by vertical lines which correspond
to the $t^{*}\leq\mathcal{L}_{PAD}^{Ising}$ (i.e. the Ising like
PAD extension) (orange line), the $S_{2}\sqrt{t_{0}^{*}}=1$ condition
(black line), the $t^{*}=t_{\gamma}^{*}$ condition {[}Eq. (\ref{eP 1d2 (25)}){]}
(blue line), and $t^{*}\geq\mathcal{L}_{PAD}^{mf}$ (i.e. the mean
field like PAD extension) (pink line), respectively (see text for
details). The amplitude and propagation of the asymptotic theoretical
error-bars for the non-gausssian limit $t^{*}\rightarrow0$ are clearly
evident in the left part. To compare the amplitude of the theoretical
error-bar to the one of the experimental uncertainty, the squared
point and its associated theoretical error-bar indicate {}``equivalence''
between the model and the experimental situation encountered at the
{}``lowest '' temperature distance $T-T_{c}=1\, mK$ above the critical
temperature of xenon (see Ref. \cite{Garrabos2006a} for details).
The rigth part corresponds to the residual magnification at the scale
of the allowed \cite{Bagnuls2002} theoretical precision level ($\leq0.01$\%).
	\label{Figure 5}}
\end{figure*}

\begin{figure*}
	\includegraphics[width=2\columnwidth]{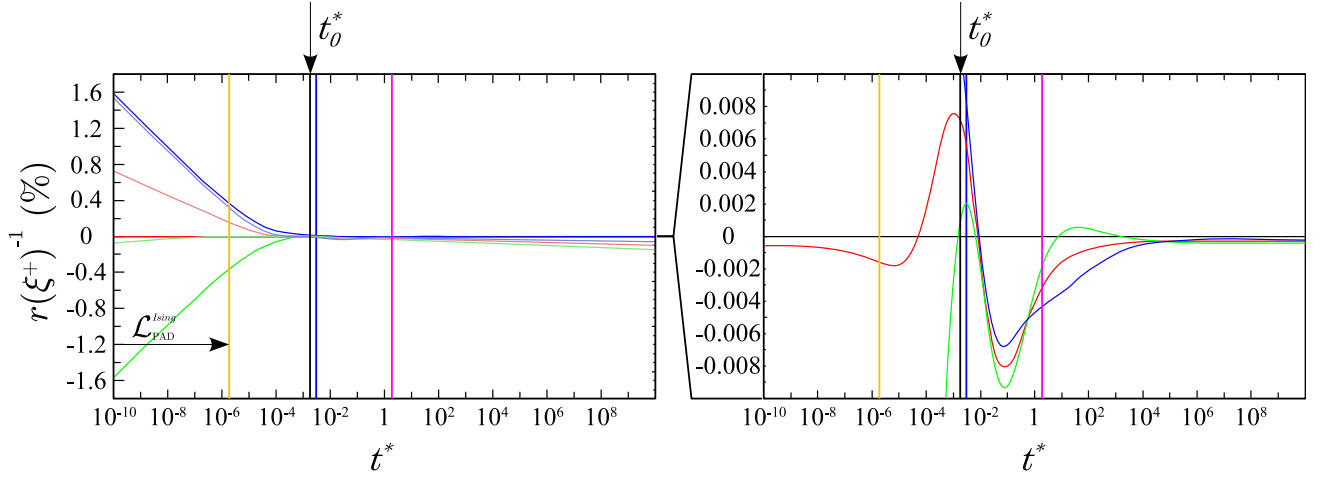}
	\caption{Same as Figure 4 for the inverse correlation length in the homogeneous
phase.
	\label{Figure 6}}
\end{figure*}

\begin{figure*}
	\includegraphics[width=2\columnwidth]{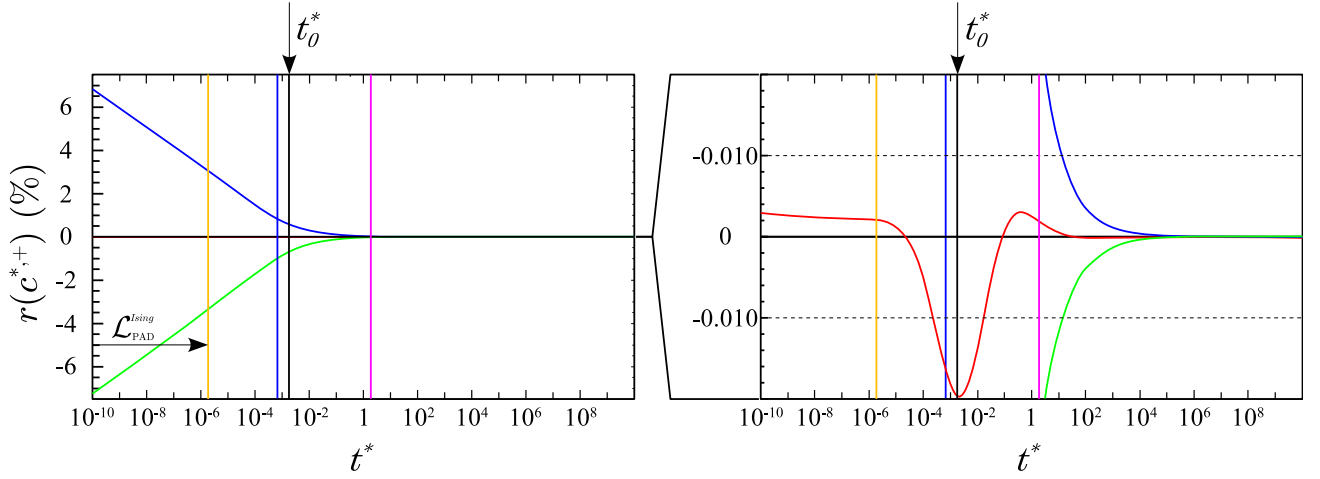}
	\caption{Same as Figure 4 for the specific heat in the homogenous phase. Compared
to results of Figures 4 and 5, the significative increases in amplitude
and propagation of the error-bars is such as the residuals attain
the admitted theoretical level in the C3 domain.
	\label{Figure 7}}
\end{figure*}

\begin{figure*}
	\includegraphics[width=2\columnwidth]{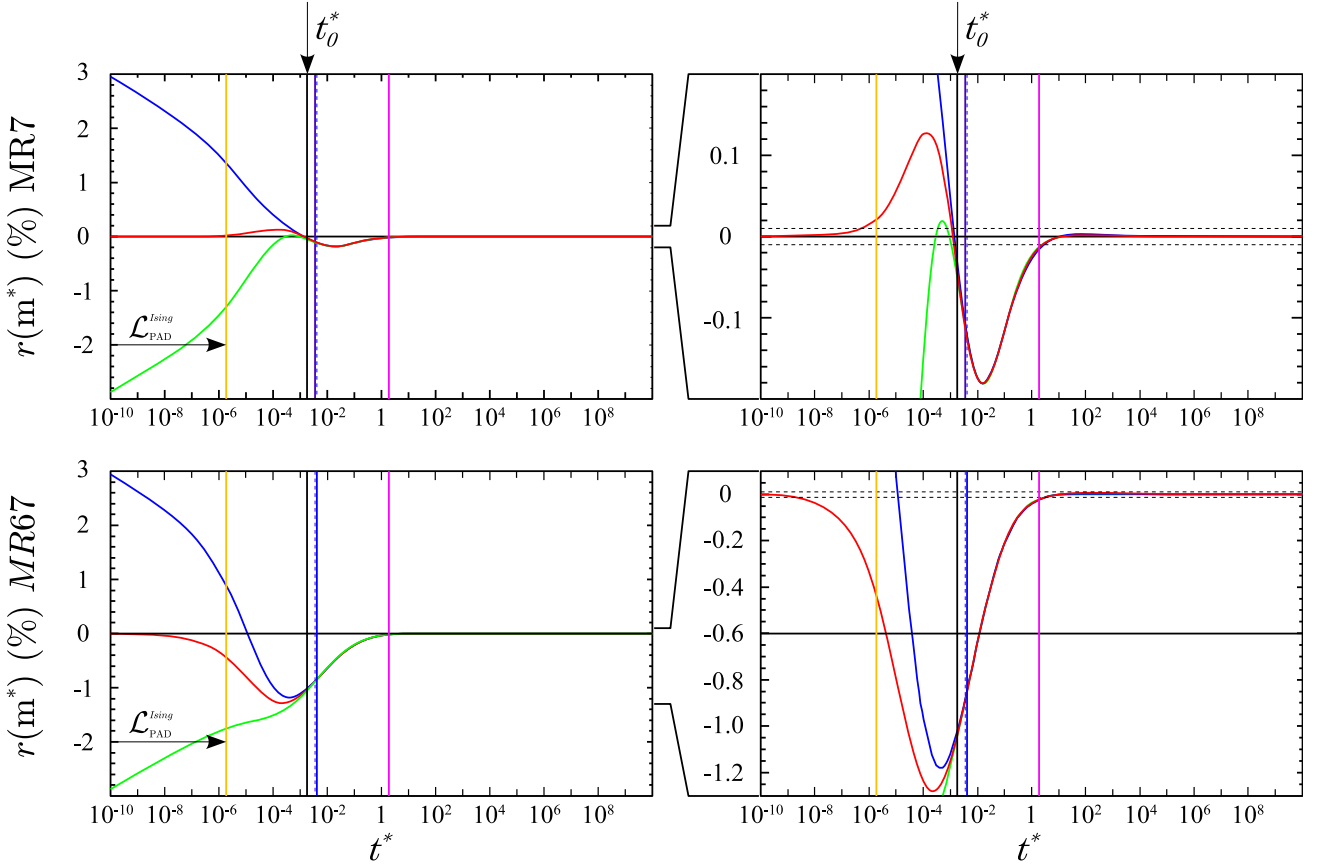}
	\caption{Residuals (\%) for the order parameter in the non-homogenous phase.
See Appendix B for details between the MR7 (upper part) and MR 67
(lower part) results.
	\label{Figure 8}}
\end{figure*}

\begin{figure*}
	\includegraphics[width=2\columnwidth]{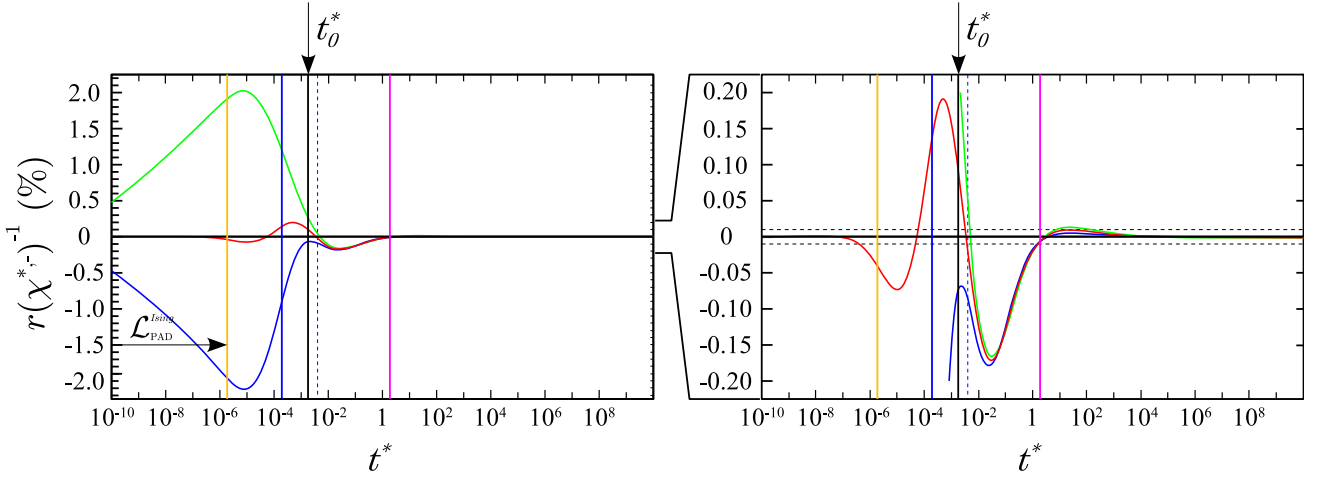}
	\caption{Same as Figure 4 for the susceptibility in the non-homogenous phase.
Compared to the general trend of all the other max and min crossover
functions, a curious {}``exact crossing'' between min and max functions
occurs in the present case around $t^{*}\simeq3\,10^{-13}$, as revealed
by the converging residuals for $t^{*}<10^{-5}$. That provides the
maximun amplitude for theoretical error-bars observed outside the
PAD extension. The associated error-bar propagation is then significative
in a large C3 domain.
	\label{Figure 9}}
\end{figure*}

\begin{figure*}
	\includegraphics[width=2\columnwidth]{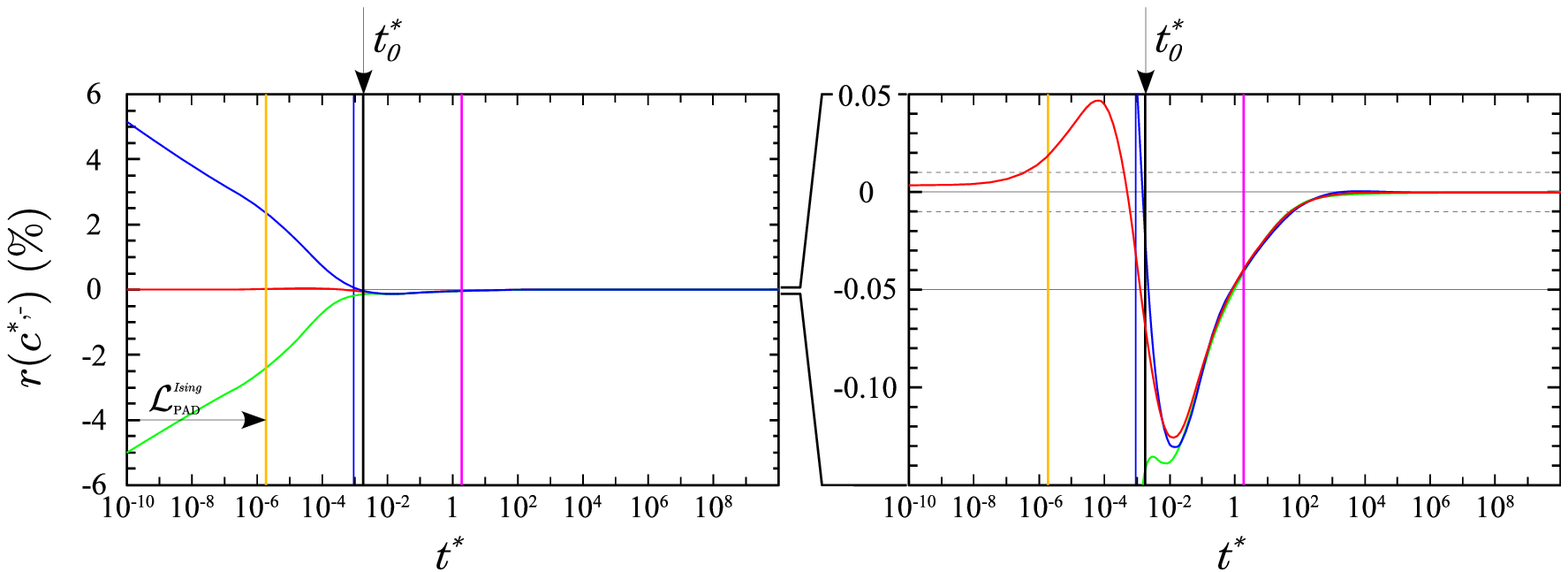}
	\caption{Same as Figure 4 for the specific heat in the non-homogenous phase.
Compared to the homogeneous phase (Figure 6), we suspect that the
non-zero value of the backgroung term contributes to increase amplitude
and error-bar propagation over the C3 domain, leading to its over-extended
range close to the Gaussian limit $t^{*}\rightarrow\infty$.
	\label{Figure 10}}
\end{figure*}

In Figure \ref{Figure 4}, using a vertical
arbitrary unit $(a.u.)$ scale between $e_{Pmin,\frac{1}{2}}(a.u.)=0$
to $e_{Pmax,\frac{1}{2}}(a.u.)=1$, then with $e_{P,\frac{1}{2}}(a.u.)=e_{Pmix,\frac{1}{2}}(a.u.)=\frac{1}{2}$
at $t^{\ast}=t_{e_{P}}^{\ast}$, permits to generalize the (confluent)
$\Delta$ case reported in Figure \ref{Figure 3}. We can observe
that the conditions of Eq. (\ref{eP 1d2 (25)}) match perfectly at
$t^{\ast}=t_{e_{P}}^{\ast}$ for mean crossover functions and mixing
crossover functions.

Figure \ref{Figure 4} is the second result supporting our $S_{2}=cte$
suggestion to construct the {}``universal'' confluent function.
The corresponding condition $\Delta_{\frac{1}{2}}=\frac{\Delta+\Delta_{mf}}{2}$
for the effective confluent exponent, where $S_{2}\sqrt{t_{0}^{\ast}}=1$
{[}Eq. (\ref{tstardelta confluent condition vs S2 (39)}){]}, demonstrates
that the $t_{e_{P}}^{\ast}$-values are readily {}``C3'' in nature
(see Figure \ref{Figure 3}), since, either the conditions $\frac{1}{10}\left(\frac{1}{S_{2}}\right)^{2}\lesssim t_{e_{P}}^{\ast}\lesssim2\left(\frac{1}{S_{2}}\right)^{2}$,
or the conditions $\left|\Delta_{eff}\left(t_{e_{P}}^{\ast}\right)-\Delta_{\frac{1}{2}}\right|<\frac{1}{2}\left|\Delta_{Asymp}-\Delta_{\frac{1}{2}}\right|$,
with $\Delta_{Asymp}=\Delta\, or\,\Delta_{mf}$, are satisfied. We
note that the relative $t^{\ast}$-differences between the values
of $t_{\gamma}^{\ast}\left(-\right)\approx\frac{1}{10}\left(\frac{1}{S_{2}}\right)^{2}$,
and $t_{\gamma}^{\ast}\left(+\right)\approx2\left(\frac{1}{S_{2}}\right)^{2}$,
for the inverse susceptibility in the non-homogeneous $\left(-\right)$,
and homogeneous $\left(+\right)$, phases, is correctly reproduced,
while the $t^{\ast}$-similarity, $t_{\alpha}^{\ast}\left(-\right)\approx t_{\alpha}^{\ast}\left(+\right)\approx\frac{1}{3}\left(\frac{1}{S_{2}}\right)^{2}$
of the corresponding values for the specific heat, is also recovered
(compare with the Figure 1 of Ref. \cite{Bagnuls2002}).

\section{Conclusion}

By construction, the mean theoretical crossover functions determined
in the present work do not account for any uncertainty on the parameters
which, in the preasymptotic critical domain (PAD), characterize the
critical behavior of the functions considered (correlation length,
susceptibility, specific heat and coexistence curve). The recourse
to such mean functions is justified by the fact that the amplitude
and propagation of the asymptotic error-bars have a large extension
which can affect significantly the classical-to-critical crossover
especially in the intermediate region where most experimental data
are reported. Actually, it is known that in such a situation, when
the comparison with experimental data is made following the common
way of trying to determine the leading and first confluent amplitudes
in the Ising-like PAD, the theoretical and experimental uncertainties
add to each other. Consequently, it is usual to fix the exponents
to their mean theoretical values in order to get some definite information
on the correction amplitudes. If we use the min and max theoretical
crossover functions very close to the critical point, we obtain non-satisfying
information on the intermediate crossover because the physical parameters
$\vartheta$ and $\psi$, introduced by the basic Eqs. (\ref{theta RG linearization (12)})
and (\ref{psy RG linearization (13)}) are not sufficiently well defined.
The mean functions constructed with controlled constraints which correctly
account for the universal features proper to the Ising-like PAD, provide
useful criteria to define the extension of the Ising-like PAD. Fitting
the experimental data with our mean crossover functions can then be
a convenient method to validate any phenomenological approach by predicting
the values of the two-scale factors which characterize the asymptotic
singular behavior within the Ising-like PAD and finally to identify
eventual subclass of universality. In a forthcoming paper \cite{Garrabos2006a},
the importance of the mean crossover functions will be illustrated
by providing a complete characterization of the one component fluid
subclass, within an extended asymptotic range well beyond the Ising
like PAD, recovering then the practical intermediate range which is
usually described by more complex formulations \cite{Agayan2001,Kiselev2003}.

\appendix

\section{The adjustable parameters}

To simplify our analysis, we select the three-dimensional Ising-like
$\Phi_{d=3}^{4}\left(n=1\right)$ model, with $g_{0}\sim\left[length\right]^{-1}$.

Let us consider the theoretical crossover function of Eq. (\ref{Function MR P (5)})
for the (dimensionless) correlation length $\ell^{*}\left(t^{\ast}\right)$
given in Table \ref{Table I} at $h^{*}=0$ as a typical example,
writing then 
\begin{equation}
	\ell^{*} = g_{0} \times \xi_{exp}
	\label{lzerostar vs gzero RG}
\end{equation}
 and {[}see Eq. (\ref{theta RG linearization (12)}){]}, 
\begin{equation}
	t^{\ast} = \vartheta \times \Delta \tau^{\ast}
	\label{tstarzero vs thetzero}
\end{equation}

Now fitting the experimental correlation length $\xi_{exp} \left( \Delta \tau^{\ast} \right)$,
in the homogeneous phase $\Delta \tau^{\ast} > 0$, gives access to the
two adjustable parameters $g_{0}$ and $\vartheta$. Consequently,
due to the asymptotic validity of Eq. (\ref{tstarzero vs thetzero})
close to the critical point, is known the asymptotic two-term Wegner
expansion of the experimental singular behavior for the correlation
length, which reads as follows \begin{equation}
\xi_{exp}=\xi_{0}^{+}\left(\Delta\tau^{\ast}\right)^{-\nu}\left[1+a_{\xi}^{+}\left(\Delta\tau^{\ast}\right)^{\Delta}+...\right]\label{two term correlation length}\end{equation}
 The associated two-term expansion of the theoretical function reads
as follows

\begin{equation}
\begin{array}{c}
\ell^{*}=\left(\mathbb{Z}_{\ell}^{+}\right)^{-1}\left(t^{\ast}\right)^{-\nu}\left[1-\mathbb{Z}_{\ell}^{1,+}\left(t^{\ast}\right)^{\Delta}+...\right]\end{array}\label{two term theoretical inverse corr length}\end{equation}
 Using Eqs. (\ref{lzerostar vs gzero RG},\ref{tstarzero vs thetzero}),
a term-to-term identification between Eqs. (\ref{two term correlation length})
and (\ref{two term theoretical inverse corr length}), gives,

\begin{equation}
\vartheta=\left(\frac{a_{\xi}^{+}}{-\mathbb{Z}_{\ell}^{1,+}}\right)^{\frac{1}{\Delta}}\label{thetazero RG vs correlation length}\end{equation}
 and \begin{equation}
g_{0}^{-1}=\xi_{0}^{+}\mathbb{Z}_{\ell}^{+}\vartheta^{\nu}\label{gzero RG vs correlation length}\end{equation}
 Within the preasymptotic domain, the two adjustable parameters of
the model at $h^{*}=0$ are unequivocally defined by Eqs. (\ref{thetazero RG vs correlation length})
and (\ref{gzero RG vs correlation length}). Two important remarks
can be made:

i) the nonuniversal scale factor $\vartheta$ is uniquely defined
by the confluent corrections to scaling;

ii) the nonuniversal inverse length $g_{0}$ is proportional to the
inverse of the leading amplitude of the actual correlation length.

>From now on, the complete characterization of the system would follow
from the determination of $\psi$ which requires the consideration
of one suplementary singular property, such as the susceptibility
(as shown in the second paper \cite{Garrabos2006a}).

However, the inverse length $g_{0}$, determined from correlation
length measurements, is not a {}``natural'' scale for the actual
system because a measurement of the correlation length would be a
prerequisite condition to set the length scale unit. One may easily
show \cite{Privman1991} that when the thermodynamic description of
a magnetic system is normalized per particle, then all lengths are
measured in units of the thermodynamic microscopic length $a_{mag}=(v_{mag})^{1/d}$
(where $v_{mag}$ is the particle volume at criticality). In this
respect, $a_{mag}$ plays a role similar to that of the lattice spacing
$a_{Ising}$ of the Ising uncompressible solid. The natural approach
is thus to first choose a microscopic length unit such as $a_{Ising}$,
which implicitly introduces the number of particles per lattice cell.
Therefore, the value of $g_{0}$ is obtained via the length $a_{Ising}$
from the determination of an intermediate parameter $u_{Ising}^{*}$
{[}like in Eq. (\ref{microlengthscale (15)}){]} such as: \begin{equation}
g_{0}\times a_{Ising}=u_{Ising}^{*}\label{uIsing}\end{equation}

As a conclusion, fitting experimental results with the dimensionless
theoretical functions requires reference to one length scale unit
{[}$a_{Ising}$, or $a_{mag}${]} and the energy scale unit $k_{B}T_{c}$
to reduce to dimensionless quantities the thermodynamic and correlation
functions. The fitting results of correlation length and susceptibility
measurements for example \cite{Garrabos2006a}, enable the determination
of three dimensionless numbers $\vartheta$, $\psi$, and $g_{0}\times a_{Ising}$
(or $g_{0}\times a_{mag}$). In such a situation, $\vartheta$ and
$\psi$ act as the two-scale factors characteristic of the asymptotic
universal features of the Ising-like system, while the product $g_{0}\times a_{Ising}$
(or $g_{0}\times a_{mag}$) insures that the extensivity of the total
macroscopic system is correctly accounted for in units of the actual
critical correlation length.

However, to compare between two systems of the same universality class
remains not easy in the absence of explicit thermodynamic definition
of the (coupling constant) inverse length $g_{0}$. This exercice
is left to the second paper \cite{Garrabos2006a}.

\section{The coexistence curve and its confluent correction error-bar}

As explained in Ref. \cite{Bagnuls2002}, the MR7 crossover functions
involve a forced account of the Guida and Zinn-Justin estimates \cite{Guida1998}
of the universal combinations between the \emph{leading} critical
amplitudes. This has induced an overestimation of the uncertainty
on the correction terms in MR7 compared to the previous MR6 calculations
(see section IIB2 of Ref. \cite{Bagnuls2002}). In general this does
not have an important impact on the resulting mean crossover functions,
except for the order parameter for which the MR7 determination finally
appears to have (relatively to MR6) a poor quality (one may appreciate
this difference by looking at bottom of Figures 2 (MR6) and 3 (MR7)
of Ref. \cite{Zhong2004}, for example).

To circumvent this non-satisfactory error-bar situation, we have applied
the procedure described in Section 3 to construct a {}``modified''
crossover function (labelled MR67) of the order parameter. Actually,
our MR67 function incorporates the same mean value of the leading
amplitude as the MR7 one, $\mathbb{Z}_{M}\left(MR67\right)=\mathbb{Z}_{M}\left(MR7\right)=0.937528$,
and combines the MR6 central value of the universal ratio $\frac{\mathbb{Z}_{M}^{1}}{\mathbb{Z}_{\chi}^{1,+}}\left(MR6\right)=0.9$,
with the MR7 amplitude value of $\mathbb{Z}_{\chi}^{1,+}\left(MR7\right)=8.5635$,
so that $\mathbb{Z}_{M}^{1}\left(MR67\right)=7.70712$. Using such
a practice, we have reported the difficulty of accounting for error
bars in the MR7 calculations of the crossover functions on only one
property in the non homogeneous range. In the forthcoming paper dedicated
to the study of the one component fluid subclass \cite{Garrabos2006a},
a detailed analysis of this particular choice will be presented.

The corresponding numerical values of the parameters of Eq. (\ref{Function MR P (5)}),
needed to calculate the MR67 mean crossover function for the order
parameter in non-homogeneous state ($t^{\ast}<0$), are given in Table
\ref{Table II}.

For a reader interested in checking the level of precision for the
present mean crossover functions in the three Ising-like, intermediate,
and mean-field like, $t^{\ast}$-range, we have reported the residuals
$100\times\left(\frac{F_{P,max,min,mix}\left[t^{\ast}\right]}{F_{P,mean}\left(t^{\ast}\right)}-1\right)$
for all the properties (see Figures \ref{Figure 5} to \ref{Figure 10}).
The right part of each figures gives the appropriate magnified scale
of the residuals to compare with the estimated precison ($\pm0.01$\%)
for the fitting theoretical functions.

\end{document}